\documentclass[12pt]{JHEP3}
\usepackage{graphicx,amsmath,amssymb}
\newcommand{\lta}{{\mathbin{\lower 3pt\hbox
   {$\rlap{\raise 5pt\hbox{$\char'074$}}\mathchar"7218$}}}} 
\newcommand{\gta}{{\mathbin{\lower 3pt\hbox
   {$\rlap{\raise 5pt\hbox{$\char'076$}}\mathchar"7218$}}}} 
\newcommand{\msun}{{M_{\odot}}}
\newcommand{\be}{\begin{equation}}
\newcommand{\ee}{\end{equation}}
\newcommand{\ba}{\begin{eqnarray}}
\newcommand{\ea}{\end{eqnarray}}
\newcommand{\dndldtfrac}{\frac{dn}{dtd \linvariant}}
\newcommand{\linvariant}{l}
\newcommand{\linvariantborn}{\linvariant_{b}}
\newcommand{\tinvariantborn}{t_{b}}
\newcommand{\dndlfrac}{\frac{dn}{d\linvariant}} 
\newcommand{\dndl}{dn/d\linvariant}
\newcommand{\dndlogl}{dn/d\log \linvariant}
\newcommand{\lgravcutoff}{\linvariant_g}
\newcommand{\mgravcutoff}{{\rm M}_g}
\newcommand{\D}{\bar D}
\def\SECTIONPAD{}
\newcommand{\pad}{\ifdefined\SECTIONPAD\newpage\fi}

\dedicated{\centering This article is to be published in International Journal of Modern Physics D and also in the book ''One Hundred Years of General Relativity: From Genesis and Empirical Foundations to Gravitational Waves, Cosmology and Quantum Gravity'', edited by Wei-Tou Ni (World Scientific, Singapore, 2015).}

\title{Inflation, String Theory and Cosmic Strings}

\author{David F. Chernoff\\ Department of Astronomy, Cornell University, Ithaca, New York}
\author{S.-H. Henry Tye\\ Jockey Club Institute for Advanced Study and Department of Physics, Hong Kong University of Science and Technology, Clear Water Bay, Hong Kong and
\\Department of Physics, Cornell University, Ithaca, New York}

\abstract {At its very beginning, the universe is believed to have
grown exponentially in size via the mechanism of inflation.
The almost scale-invariant density perturbation spectrum
predicted by inflation is strongly supported by cosmological observations, in particular the
cosmic microwave background radiation.
However, the universe's precise inflationary scenario remains
a profound problem for cosmology and for fundamental
physics.  String theory, the most-studied theory as the final physical theory of nature,
should provide an answer to this question. Some of the proposals on how inflation is realized in string theory are reviewed. 
 Since everything is made of strings, some string loops of cosmological sizes are likely to survive in the hot big bang that followed inflation.
They appear as cosmic strings, which can have intricate properties. Because of the warped geometry in flux compactification of the extra spatial dimensions in string theory, some of the cosmic strings may have tensions substantially below the Planck or string scale.
Such strings cluster in a manner similar to dark matter leading to
hugely enhanced densities.
As a result, numerous fossil remnants of the low tension cosmic
strings may exist within the galaxy. They can be revealed
through the optical lensing of background stars in the near future and
studied in detail through gravitational
wave emission. We anticipate that these cosmic strings will permit us to address central questions about the properties of string theory as well as the birth of
our universe.}

\begin {document}

\section{Introduction}
By 1930, quantum mechanics, general relativity and the Hubble expansion of our universe were generally accepted by the scientific community as 3 triumphs of modern physical science. Particles and fields obey quantum mechanical rules and space-time bends and warps according to Einstein's classical description of gravity. The dynamical arena for particles, fields and space-time is the universe and modern cosmology is born as the application of these physical laws to the universe itself. 

The path forward seemed clear but in the 1950s and 1960s physicists first began to appreciate that trying to quantize the classical gravitational field led to severe inconsistencies. This non-renormalizability problem was further sharpened once the quantization of gauge (vector) fields was well understood in early 1970s. 

At about the same time, string theory -- a quantum mechanical theory of one-dimensional objects -- was found to contain a graviton (a massless spin-2 particle). The promise of a self-consistent theory of quantum gravity has attracted huge attention since the 1980s. Tremendous progress has been made in the past 30 years in understanding many aspects of string theory; yet today we are no closer to knowing which of many theoretical possibilities might describe our universe and we still lack experimental evidence that the theory describes nature. 

By now, string theory is such a large topic with so many research directions that we have to choose a specific area to discuss. The centenary of general relativity prompts us to reflect on 2 very amazing manifestations of general relativity: namely black holes and cosmology. Given the tremendous progress in cosmology in the past decades, both in observation and in theory, we shall focus here on the intersection of string theory and cosmology. 

Our fundamental understanding of low energy physics is grounded in quantum field theory and gravity but if we hope to work at the highest energies then we cannot escape the need for an ultraviolet completion to these theories. String theory provides a suitable framework. The string scale, $M_S=1/\sqrt{\alpha'}$ where $\alpha'$ is the Regge slope, is expected to be below the reduced Planck scale $M_{Pl} = 1/\sqrt{8 \pi G} \simeq 2.4 \times 10^{18}$ GeV (where $G$ is the Newton's constant) but not too far from the grand unified scale $M_{GUT} \simeq 10^{16}$ GeV (where the proton mass is about 1 GeV, in units where $c=\hbar  =1$). Particle physics properties at scales much lower than $M_S$ are hard to calculate within our present understanding of string theory. High energy experiments can reach only multi-TeV scales, orders of magnitude below the expected $M_S$. The difficulty of finding a common arena to compare theory and observation is hardly a new dilemma: although standard quantum mechanics works well at and above eV energy scales, it is woefully inadequate to describe properties of proteins which typically involve milli-eV scales and lower. Where should we look?

One place to look for stringy effects is in the early universe. If the energy scale of the early universe reaches $M_S$, an abundance of physical effects will provide a more direct path to test our ideas than currently feasible in high energy experiments. In fact, attempts to interpret observations will tell us whether our present understanding of physics and the universe is advanced enough to permit us to ask and answer sensible questions about these remote physical regimes. We shall focus on the inflationary universe scenario
\cite{guth_inflationary_1981,linde_new_1982,albrecht_cosmology_1982}, 
which reaches energy scales sufficiently high to create ample remnants. The hot big bang's nucleosynthetic epoch has direct observational support at and below MeV scales whereas the energies of interest to us are $\sim M_S$. We are most fortunate that precision cosmology has begun to let us address questions relevant to these extreme conditions. Even in this rather specialized topic of inflationary cosmology and string theory, a comprehensive review by Baumann and McAllister \cite{Baumann:2014nda} appeared recently. So we shall take the opportunity here to present a brief introduction and share within this framework some of our own views on a particularly interesting topic, namely, cosmic strings. The subject of cosmic strings has been extensively studied too 
\cite{vilenkin_cosmic_2000},
so here we shall focus on low tension cosmic strings, which may appear naturally in string theory.

Since current gravitational observations are insensitive to quantum effects, one may wonder why an ultraviolet completion of the gravity theory is needed if inflation, in fact, occurs several orders of magnitude below the Planck scale. To partly answer this, 
we like to recall that in 1920s, there was no obvious necessity for a quantum theory consistent with special relativity given that electrons in atoms and materials move at non-relativistic speed only. However, the Dirac theory predicted anti-matter, the value of the electron's  gyromagnetic moment and a host of interesting properties (e.g., the Lamb shift) that are well within reach of the non-relativistic study of electrons and atoms. The standard electroweak model, the ultraviolet completion of the Fermi's weak interaction model, is another example. Among other successes, the standard electroweak model predicted the presence of the neutral current as well as explains the $\pi^0$ decay to 2 photons via anomaly and the basis of the quark-lepton family via the anomaly free constraints.
All these phenomena are manifest at energy scales far below the electroweak scale.
In the current context, we need to search for stringy effects that are unexpected in classical general relativity and might be observed in the near future at energy scales far below the Planck scale.

There are many specific ways to realize inflation within string theory. These may be roughly grouped into models with small and large field ranges. They have rather different properties and predictions. Recall that 6 of the 9 spatial dimensions must be compactified into a Calabi-Yau like manifold with volume $V_6$, where
\be
\label{V6}
\left(\frac{M_{Pl}}{M_S}\right)^2 \sim M_S^6V_6 \gg 1 .
\ee
During inflation the inflaton field $\phi$ moves from an initial to final position within the manifold, covering a path or field range $\Delta \phi$. Now, the distance scale inside the compactified manifold is bounded by $M_{Pl}$. The case $\Delta \phi < M_{Pl}$ is known as small field inflation. The opposite limit when the field range far exceeds the typical scale of the manifold, $\Delta \phi \gg M_{Pl}$, is large field inflation. In the latter case the inflaton wanders inside the compactified manifold for a while during inflation. This scenario may occur if the inflaton is an axion-like field and executes a helical-like motion. Flux compactification of the 6 spatial dimensions in string theory typically introduces many axions and naturally sets the stage for this possibility.

The recent B-mode polarization measurement in the cosmic microwave background (CMB) radiation
\cite{Ade:2014xna}
by BICEP2 is  most interesting. Primordial B-mode polarization is sourced by tensor perturbations during inflation and measurable primordial B-mode polarization implies that $\Delta \phi \gg M_{Pl}$. This is a prototypical example of how it's now possible to probe high energies and early moments of the universe's history by means of cosmological observations. However, since dust can contaminate today's observed signal \cite{Flauger:2014qra,Mortonson:2014bja}, it remains to be seen whether the BICEP2 observations definitively imply that large field inflation takes place. We believe this issue should be settled soon. In the meantime, we shall entertain both possibilities here.

In both small and large field ranges macroscopic, one-dimensional objects, hereafter cosmic strings, can appear rather naturally. In the former case, the cosmic strings can be the fundamental superstrings themselves, while in the latter case, they can be fundamental strings and/or vortices resulting from the Higgs like mechanism. Cosmic strings with tension above the inflation scale will not be produced after inflation and we should expect only relatively low tension strings to be produced. Because of the warped geometry and the presence of throats in the compactification in string theory, some cosmic strings can acquire very low tensions. In flux compactification such strings tend to have nontrivial tension spectrum (and maybe even with beads).

If the cosmic string tension is close to the present observational bound (about $G\mu <10^{-7}$), strings might contribute to the B-mode polarization at large $\ell$ multipoles independent of whether or not BICEP2 has detected primordial B mode signals. The string contribution to the CMB power spectrum of temperature fluctuations is limited to be no more than $\sim 10$\% and this constrains any string B-mode contribution. Because the primordial tensor perturbation from inflation decreases quickly for large $\ell$ multipoles one very informative possibility is that cosmic strings are detected at large $\ell$ and other astrophysical contributions are sub-dominant.

On the other hand, if tension is low ($G\mu \ll10^{-7}$) then the cosmic string effect for all $\ell$ is negligible compared to known sources and foreground contributions. Other signatures must be sought. For small $G\mu$ string loops are long lived and the oldest (and smallest) tend to cluster in our galaxy, resulting in an enhancement of $\sim 10^5$ in the local string density. This enhancement opens up particularly promising avenues for detecting strings by microlensing of stars within the galaxy and by gravitational wave emission in the tension range $10^{-14} < G\mu \ll10^{-7}$. As a cosmic string passes in front of a star, its brightness typically doubles, as the 2 images cannot be resolved. As a string loop oscillates in front of a star, it generates a unique signature of repeated achromatic brightness doubling. Here we briefly review the various observational bounds on cosmic strings and estimate the low tension cosmic string density within our galaxy as well as the likelihood of their detection in the upcoming observational searches. It is encouraging that searches of extrasolar planets and variables stars also offer a chance to detect the micro-lensing of stars by cosmic strings. Once a location is identified by such a detection, a search for gravitational wave signals should follow.

Detection of cosmic strings followed by the measurement of their possible different tensions will go a long way in probing superstring theory. Although a single string tension can easily originate from standard field theory, a string tension spectrum should be considered as a distinct signature of string theory. It is even possible that some cosmic strings will move in the compactified dimensions with warped geometry, which can show up observationally as strings with varying tension, both along their lengths as well as in time. In summary, cosmic strings probably offer the best chance of finding distinct observational support for the string theory.

\pad

\section{The Inflationary Universe}

So far, observational data agrees well with the simplest version of the slow-roll inflationary universe scenario, i.e., a single, almost homogeneous and isotropic, scalar inflaton field $\phi=\phi(t)$ subject to potential $V(\phi)$.
For a Friedmann-Lema\^{i}tre-Robertson-Walker metric, general relativity yields schematic simple equations for the cosmic scale factor $a(t)$ and for $\phi$,
\ba
\label{GRINF}
H^2 = \left(\frac{\dot a}{a}\right)^2 &=& \frac{1}{3 M_{Pl}^2} [V(\phi)+ \frac{{\dot \phi}^2}{2}] + \frac{\rho_c(0)}{a^2} + \frac{\rho_m(0)}{a^3}+ \frac{\rho_r(0)}{a^4} + . . .\\
\label{GRINF2}
\ddot \phi +3H \dot \phi &=& - \frac{dV}{d\phi} = - V'(\phi)
\ea
where the Hubble parameter $H=\dot a/a$ ($\dot a(t) =da/dt$) measures the rate of expansion of the universe, and $\rho_c(0)$, $\rho_m(0)$
and $\rho_r(0)$ are the curvature, the non-relativistic matter and the radiation densities at time $t=0$.\footnote{The energy density has been divided into $\rho_m$ and $\rho_r$. The ultra-relativistic massive particles are lumped into $\rho_r$ and the marginally relativistic massive particles included in $\rho_m$. This division is epoch-dependent and schematic.} In an expanding universe, $a(t)$ grows and the curvature, the matter and the radiation terms diminish. The energy and pressure densities for the inflaton are
\ba
\rho_{\phi} =  \frac{{\dot \phi}^2}{2} + V(\phi) \\\nonumber
p_{\phi} = \frac{{\dot \phi}^2}{2} - V(\phi)
\ea
where a canonical kinetic term for $\phi$ is assumed.
If $V(\phi)$ is sufficiently flat and if $\dot \phi$ is initially small then Eq.(\ref{GRINF2}) implies that $\phi$ moves slowly in the sense that its kinetic energy remains small compared to the potential energy, ${\dot \phi}^2 \ll V(\phi)$. In the limit that $V(\phi)$ is exactly constant and dominates in Eq.(\ref{GRINF}) then $H$ is constant and $a(t) \propto e^{Ht}$. More precisely one may define the inflationary epoch as that period when the expansion of the universe is accelerating, i.e., $\ddot a >0$, or, in a spatially flat universe, 
\ba
2 M_{Pl}^2\dot H &=& -(\rho + p) \\\nonumber
\epsilon & = & -\frac{\dot H}{H^2} =  \frac{3}{2} \left( 1 + \frac{p}{\rho} \right) \\\nonumber
         & < & 1 .
\ea
Slow-roll inflation means small $\epsilon$, $H$ nearly constant and $a(t) \sim a(0) e^{Ht}$.

In a typical slow-roll inflationary model inflation ends at $t=t_{end}$ when the inflaton encounters a steeper part of the potential and $\epsilon > 1$. The number of e-folds of inflation is $N_e \simeq Ht_{end}$, a key parameter. The energy released from the potential $V(\phi)$ heats the universe at the end of inflation and starts the hot big bang. The period of slow-roll must last at least $N_e > 50$ e-folds to explain three important observations about our universe that are otherwise unaccounted for in the normal big bang cosmology: flatness, lack of defects and homogeneity. For any reasonable initial curvature density $\rho_c(0)$, the final curvature density $\rho_{c}(t_{end}) < \rho_c(0)e^{-150}$ will be totally negligible, thus yielding a flat universe. This is how inflation solves the flatness problem. Any defect density (probably included in $\rho_m(0)$) present in the universe before the inflationary epoch will also be inflated away, thus solving the so-called ``monopole'' or defect problem. Since the cosmic scale factor grows by a huge factor, the universe we inhabit today came from a tiny patch of the universe before inflation. Any original inhomogeneity will be inflated away. Inflation explains the high degree of homogeneity of the universe. In summary, if $N_e$ is sufficiently large then inflation accounts for the universe's observed flatness, defect density and homogeneity. 

What is amazing is that inflation also automatically provides a mechanism to create primordial inhomogeneities that ultimately lead to structure formation in our universe.
As the inflaton slowly rolls down the potential in the classical sense, quantum fluctuations yield slightly different ending times $t_{end}$ and so slightly different densities in different regions. The scalar and the metric fluctuations may be treated perturbatively, and one obtains the dimensionless power spectra in terms of $H$,
\ba
\Delta_S^2 (k) &=& \frac{1}{8 \pi^2}\frac{H^2}{M_{Pl}^2|\epsilon|} \\\nonumber
\Delta_T^2 (k) &=& \frac{2}{\pi^2}\frac{H^2}{M_{Pl}^2}= r\Delta_S^2 (k) \\\nonumber
r&=&16 \epsilon
\ea
where $\Delta_S^2$ and $\Delta_T^2$ are the scalar and the tensor modes respectively. A scalar scale-invariant power spectrum corresponds to constant $\Delta_S^2$, which occurs when a space expands in a nearly de-Sitter fashion for a finite length of time.
The scalar mode is related to the temperature fluctuations first measured by COBE \cite{smoot_structure_1992}. Measurements and modeling have been refined over the past 2 decades.
Since $\phi$ rolls in the non-flat potential in the inflationary scenario, $\Delta_S^2$ will have a slight $k$ dependence. This is usually parametrized with respect to the pivot wave number $k_p$ in the form
\ba
\Delta_S^2 (k) &=& \Delta_S^2(k_p) (k/k_p)^{n_s -1 + (dn_s/d\ln k) \ln(k/k_p)/2 + . . . } \\
\Delta_T^2 (k) &=& \Delta_T^2 (k_p) (k/k_p)^{n_t +  (dn_t/d\ln k) \ln(k/k_p)/2 + . . . } 
\ea
where the PLANCK and WMAP best fit values for $\Lambda$CDM
quoted in Ref.\cite{Ade:2013uln} for $k_p = 0.05$Mpc$^{-1}$ are
\ba 
n_s & = & 0.9603 \pm 0.0073 \\
\Delta_S^2(k_p) & = & 2.19^{+0.58}_{-0.53} \times 10^{-9} .
\ea
These do not change appreciably when a possible tensor component is
included and PLANCK constraints on $r$ are given at pivot $0.002$ Mpc$^{-1}$.
In the slow-roll approximation, where $\ddot \phi$ is negligible in Eq.(\ref{GRINF2}), the deviation from scale-invariance is quantified by the spectral tilt 
\ba
n_s-1&=& \frac{d \ln \Delta_S^2 (k) }{d \ln k} = -6 \epsilon +2 \eta \\\nonumber
 { dn_s \over d \ln k} &=& -16\epsilon\eta + 24 \epsilon^2 +2\xi^2  
 \ea
 where the parameters that measure the deviation from flatness of the potential are
 \ba
\epsilon &=& \frac{M_{Pl}^2}{2} (V'/V)^2,  \\
\eta & = & M_{Pl}^2V''/V, \\\nonumber
\xi^2 &=& M_{Pl}^4V'V'''/V^2 . 
\ea
where $\epsilon$ and $\eta$ are the known as the slow-roll parameters which take small values during the inflationary epoch.
The tensor mode comes from the quantum fluctuation of the gravitational wave. Its corresponding tilt is
\be
n_t = - 2 \epsilon .
\ee

For a small field range $\Delta \phi$, the constraint on $N_e$ dictates a rather flat potential and small $\epsilon$, which in turn implies a very small $r$.
This relation can be quantified by the Lyth bound,
 \cite{Lyth:1996im},
\begin{equation}
\label{Lyth}
 \frac{\Delta \phi}{M_{Pl}} \ge  N_{e} \sqrt{r/8} .
\end{equation}
If $60 \ge N_e \ge 40$, we find that typical values of $r$ satisfies $r < 0.005$ for $\Delta \phi < M_{Pl}$. This is much smaller than $r \simeq 0.2$ reported by BICEP2 \cite{Ade:2014xna}. 

While $\Delta_S^2$ is a combination of $H^2$ and $1/\epsilon$, $\Delta_T^2$ provides a direct measurement of $H^2$ and hence the magnitude of the inflaton potential during inflation:
\be
\label{Vr-relation}
V \simeq \left( \frac{r}{0.2} \right) \left( 2.2 \times 10^{16}  {\rm GeV} \right)^4 .
\ee
The imprint of tensor fluctuations are present in the CMB but $\Delta_T^2 \ll \Delta_S^2$ so it's not possible to measure $\Delta_T^2$ directly from total temperature fluctuations at small $\ell$.\footnote{Since tensor and scalar perturbations have different $\ell$ dependencies experimentalists deduce the scalar spectrum from high $\ell$ measurements where scalar power will dominate any tensor contribution. Then they observe the total temperature fluctations at low $\ell$ where tensors should make their largest impact. If the theoretical relation between low and high $\ell$ fluctuations is known (``no running'' being the simplest possibility) then the total observed low $\ell$ power limits whatever extra contribution might arise from the tensor modes. The WMAP \cite{2013ApJS..208...19H} and Planck \cite{2014A&A...571A..16P} collaborations (incorporating data from the SPT [South Pole Telescope] and ACT [Atacama Cosmology Telescope] microwave background experiments and baryon acoustic oscillation observations) placed limits on $r$, $r<0.13$ and $r<0.11$ respectively. These upper limits on $r$ are somewhat {\it less} than the $r$ values reported by BICEP2 and under investigation \cite{2014PhRvL.113c1301S}.} Fortunately, the CMB is linearly polarized and can be separated into E-mode and B-mode polarizations. It happens that $\Delta_T^2$ contributes to both modes while $\Delta_S^2$ contributes only to the E-mode. So a measurement of the primordial B-mode polarized CMB radiation is a direct measurement of $\Delta_T^2$. 

Searching for the B-mode CMB is very important. Besides the intrinsic smallness of the B-mode signal, which makes detection a major challenge, the primordial fluctuation may be masked by the interstellar dust. It is likely that the uncertainty from dust would have been resolved by the time this article appears. So we may consider this as a snapshot after the announcement of the BICEP2 data and before a full understanding of the impact of dust on the reported detection. Due to this uncertainty, we shall consider both a negligibly small $r$ (say $r < 0.002$) and an observable primordial $r \le 0.2$. 

\pad

\section{String Theory and Inflation}

By now, string theory is a huge research subject. So far, we have not yet identified the corner where a specific string theory solution fits what happens in nature. It is controversial to state whether we are close to finding that solution or we are still way off. By emphasizing the cosmological epoch, we hope to avoid the details but try to find generic stringy features that may show up in cosmological observations. Here we give a lightning pictorial summary of some of the key features of Type IIB superstring theory so readers can have at least a sketchy picture of how string theory may be tested. 

Because of compactification, we expect modes such as Kaluza-Klein modes to appear in the effective 4-dimensional theory. Their presence will alter the relation between the inflaton potential scale and $r$ (\ref{Vr-relation}) to :
\be
V \simeq \frac{1}{{\hat N}^2}\left( \frac{r}{0.2} \right) \left( 2.2 \times 10^{16}  {\rm GeV} \right)^4 .
\ee
where ${\hat N}$ effectively counts the number of universally coupled (at one-loop) degrees of freedom below the energy scale of interest here \cite{Antoniadis2014}. Present experimental bound on ${\hat N}$ is very loose. Surprisingly, a value as big as  ${\hat N} \sim 10^{25}$ is not ruled out.


\subsection{String Theory and Flux Compactification}

Here is a brief description on how all moduli are dynamically stabilized in flux compactification.
Recall that 2-form electric and magnetic field strengths follow from the 1-form field $A_{\mu}$ in the electromagnetic theory, under which point-like particles (such as electrons) are charged. In analogy to this, fundamental strings in string theory are charged under a 2-form field $B_{\mu \nu}$ which yields 3-form field strengths. In general, other dimensional objects are also present in string theory, which are known as branes. A p-brane spans p spatial dimensions, so a membrane is a 2-brane while a 1-brane is string like; that is a 1-brane is really a string.
In the Type IIB superstring theory, there are a special type of $D$p-branes where p is an odd integer. Among other properties of $D$p-branes, we like to mention two particularly relevant ones here : (1) each end of an open string must end on a $D$p-brane, and (2) with the presence of $D$1-string (or -brane), there exists another 2-form field $C_{\mu \nu}$ under which a $D$1-string is charged. 

Self-consistency of Type IIB superstring theory requires it to have 9 spatial dimensions.  Since only 3 of them describe our observable world, the other 6 must be compactified. Consider a stack of $D$3-branes of cosmological size. Such a stack appear as a point in the 6 compactified dimensions. In the brane world scenario,  all standard model particles (electrons, quarks, gluons, photons, et. al.) are light open string modes  whose ends can move freely inside the $D$3-branes, but cannot move outside the branes, while the graviton, being a closed string mode, can move freely in the branes as well as outside, i.e., the bulk region of the 6 compactified dimensions. In this sense, the $D$3-branes span our observable universe. We can easily replace the $D$3-branes by $D$7-branes, with 4 of their dimensions wrapping a 4-cycle inside the compactified 6-dimensional space. Dark matter may come from unknown particles inside the same stack, or from open string modes sitting in another stack sitting somewhere else in the compactified space.

As a self-consistent theory, the 6 extra dimensions must be dynamically compactified (and stabilized). This is a highly non-trivial problem. Fortunately, the 3-form field strengths of both the NS-NS field $B_{\mu \nu}$ and the R-R field $C_{\mu \nu}$ are quantized in string theory. Wrapping 3-cycles in the compactified dimensions, these fluxes contribute to the effective potential $V$ in the low energy approximation. Their presence can render the shape and the size of the compactified 6-dimensional space to be dynamically stabilized. This is known as flux compactification. Here the matter content and the forces of nature are dictated by the specific flux compactification \cite{Giddings:2001yu,Kachru:2003aw}. Warped internal space appears naturally. This warped geometry will come to play an important role in cosmology.

To describe nature, a Calabi-Yau like manifold is expected, with branes and orientifold planes. At low energy, a particular manifold may be described by a set of dynamically stabilized scalar fields, or moduli. One or more K\"ahler moduli parameterize the volume while the shape is described by the complex structure moduli, whose number may reach hundreds.  A typical flux compactification involves many moduli and three-form field strengths with quantized fluxes (see the review \cite{Douglas:2006es}). With such a large set of dynamical ingredients, we expect many possible vacuum solutions; collectively, this is the string theory landscape, or the so-called cosmic landscape. Here, a modulus is a complex scalar field. Written in polar coordinate, we shall refer to the phase degree of freedom as an axion. 

For a given Calabi-Yau like manifold, we can, at least in principle, determine the four-dimensional low energy supergravity effective potential $V$ for the vacua. To be specific, let us consider only $3$-form field strengths $F^i_3$ wrapping the three-cycles inside the manifold.
(Note that these are dual to the four-form field strengths in 4 dimensional space-time.) 
We have $V( F^i_{3}, \phi_j) \rightarrow V(n_i, \phi_j)$, ($i=1,2,...,N, \ j=1,2,..., K$)
where the flux quantization property of the $3$-form field strengths $F^i_3$ allow us to rewrite $V$ as a function of the quantized values $n_i$ of the fluxes present and $\phi_j$ are the complex moduli describing the size and shape of the compactified manifold as well as the couplings. There are barriers between different sets of flux values. For example, there is a (finite height) barrier between $n_1$ and $n_1-1$, where tunneling between $V(n_1,n_2, ..., n_N, \phi_j)$ and $V(n_1-1, n_2,..., n_N, \phi_j)$ may be achieved by brane-flux annihilation \cite{Bousso:2000xa}. 
For a given set of ${n_i}$, we can locate the meta-stable (classically stable) vacuum solutions $V(n_i, \phi_j)$ by varying $\phi_j$. We sift through these local minima which satisfy the following criteria: they have vanishingly
small vacuum energies because that is what is observed in today's universe,
and long decay lifetimes to lower energy states because our universe is
long-lived. These criteria restrict 
the manifolds, flux values and minima of interest; nonetheless, within the rather crude approximation we are studying, there still remain many solutions; and statistically, it seems that a very small cosmological constant is preferred \cite{Sumitomo:2012vx}.

\subsection{Inflation in String Theory}

When the universe was first created (say, a bubble created via tunneling from nothing), $\phi_j$ are typically not sitting at $\phi_{j, {\rm min}}$, so they tend to roll towards their stabilized values. Heavier moduli with steeper gradients probably reached their respective minima relatively quickly. The ones with less steep directions took longer. The last ones to reach their stabilized values typically would move along relatively flat directions, and they can play the role of inflatons. So the vacuum energy that drives inflation is roughly given by the potential when all moduli except the inflaton have already reached bottom.
Since the flux compactification in string theory introduces dozens or hundreds of moduli (each is a complex scalar field in the low energy effective theory approximation), one anticipates that there are many candidates for inflation. Even if we assume that just one field is responsible for the inflationary epoch in our own universe it seems likely that string theory can choose among many possibilities to realize single field inflation in many different ways.

In fact, the picture from string theory may offer additional interesting possibilities beyond field theory. For example, suppose there was a pair of brane-anti-brane present in the early universe whose tensions drive the inflation. As we shall see, the inflaton field happens to be the distance between them. After their annihilation that ends inflation, the inflaton field no longer exists as a degree of freedom in the low energy effective field theory. 

Some inflationary scenarios in string theory generate unobservably small primordial $r$ while others give observably large primordial $r$. Small field range implies small $r$, but large field range does not necessary imply large $r$, which happens when the Lyth bound (\ref{Lyth}) is not saturated. 
To obtain enough e-folds, we need a flat enough potential, so it is natural to consider an axion as the inflaton, as first proposed in natural inflation 
\cite{1990PhRvL..65.3233F,1993PhRvD..47..426A}. The axion has a natural shift symmetry $\phi \rightarrow \phi  \, +$ any constant. 
This continuous symmetry can easily be broken, via non-perturbative effects, to a discrete symmetry, resulting in a periodic potential that can drive inflation.
The typical field range is sub-Planckian implying a small $r$. We shall see shortly that there are many string theory models built out of the axion that permit large field ranges and some of these can yield large $r$.

Here, we shall describe the various string theory realizations of inflation within the framework of Type IIB string theory. The pictorially simplest models occur in the brane world scenario.  We shall present a few sample models to give a picture of the type of models that have been put forward. Readers can find a more complete list in Ref.\cite{Baumann:2014nda}. If $r$ turns out to be large, it will be interesting to see whether some of the small $r$ models can be adapted or modified to have a large $r$. For example, one can try to modify a brane inflationary scenario into a warm inflationary scenario with a relatively large $r$.

\pad
\section{Small $r$ Scenarios}

Besides fundamental superstrings, Type IIB string theory has $D$p-branes where the number of spatial dimensions p is odd. Furthermore, supersymmetry can be maintained if there are only $D$3- and $D$7-branes; so, unless specified otherwise, we shall restrict ourselves to this case. String theory has 9 spatial dimensions. Since our observable universe has only 3 spatial dimensions,  the other 6 spatial dimensions have to be compacted in a manifold with volume $V_6$. 
The resulting (i.e., dynamically derived) $M_{Pl}$ is related to the string scale $M_S$ via Eq.(\ref{V6})
and the typical field range $\Delta \phi $ is likely to be limited by $M_{Pl}>M_S$. For such a small field range, the potential has to be flat enough to allow 50 or more e-folds and the Lyth bound (\ref{Lyth}) for
$60 \ge N_e \ge 40$ implies $r < 0.005$. We shall refer to these models as small $r$ scenarios.

\subsection{Brane Inflation}

The discovery of branes in string theory demonstrated that the theory encompasses a multiplicity of higher dimensional, extended objects and not just strings.
In the brane world scenario, our visible universe lies inside a stack of $D$3-branes, or a stack of $D$7-branes.  Here, 6 of the 9 spatial dimensions are dynamically compactified while the 3 spatial dimensions of the $D$3-branes (or 3 of the $D$7-branes) are cosmologically large. The 6 small dimensions are stabilized via flux compactification \cite{Giddings:2001yu,Kachru:2003aw}; the region outside the branes is referred to as the bulk region. The presence of RR and NS-NS fluxes introduces intrinsic torsion and warped geometry, so there are regions in the bulk with warped throats (Figure  \ref{1.2}). 
Since each end of an open string must end on a brane, only closed strings are present in the bulk away from branes.

 There are numerous such solutions in string theory, some with a small positive vacuum energy (cosmological constant). 
 Presumably the standard model particles are open string modes; they can live either on $D$7-branes wrapping a 4-cycle in the bulk or (anti-)$D$3-branes at the bottom of a warped throat (Figure \ref{1.2}). The (relative) position of a brane is an open string mode. In brane inflation 
\cite{dvali_brane_1999}, one of these modes is identified as the inflaton $\phi$, while the inflaton potential is generated by the classical exchange of closed string modes including the graviton. From the open string perspective, this exchange of a closed string mode can be viewed as a quantum loop effect of the open string modes.
 
 \subsubsection*{$D$3-${\D}$3-brane Inflation}
 
In the early universe, besides all the branes that are present today, there is an extra pair of $D$3-${\D}$3-branes 
\cite{2001JHEP...07..047B,dvali_d-brane_2001}.
Due to the attractive forces present, the ${\D}$3-brane is expected to sit at the bottom of a throat. Here again, inflation takes place as the $D$3-brane moves down the throat towards the ${\D}$3-brane, with their separation distance as the inflaton, and inflation ends when they collide and annihilate each other, allowing the universe to settle down to the string vacuum state that describes our universe today. 
Although the original version encounters some fine-tuning problems, the scenario becomes substantially better as we make it more realistic with the introduction of warped geometry \cite{Kachru:2003sx}. 

\begin{figure}
\begin{center}
\includegraphics[width=9cm]{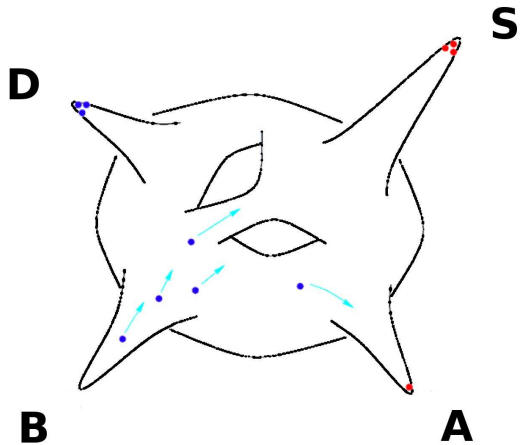}
\vspace{0.1in}
\caption{A pictorial sketch of a generic flux compactified 6-dimensional bulk, with a number of warped throats (4 of which are shown here). Besides the warped throats, there are $D$7-branes wrapping 4-cycles. 
The blue dots stand for mobile $D$3-branes while the red dots are ${\D}$3-branes sitting at the bottoms of throats. In the $D$3-${\D}$3-brane inflationary scenario, the tension of the brane pairs provide the vacuum energy that drives inflation as the $D$3-brane moves down A-throat. Inflation ends as the $D$3-brane  annihilates with the ${\D}$3-brane in $A$-throat. The standard model branes may live in A-throat or $S$-throat.
As an alternative, inflation may take place while branes are moving out of B-throat. 
}
\label{1.2}
\end{center}
\end{figure}

Because of the warped geometry, a consequence of flux compactification, a mass $M$ in the bulk becomes $h_{A}M$ at the bottom of a warped throat, where $h_{A} \ll 1$ is the warped factor (Figure \ref{1.2}). This warped geometry tends to flatten, by orders of magnitude, the inflaton potential $V(\phi)$, so the attractive $D$3-${\D}$3-brane potential is rendered exponentially weak in the warped throat. 
The attractive gravitational (plus RR) potential together with the brane tensions takes the form 
\ba
\label{infpot}
V(\phi) =  2T_3h_A^4\left(1-\frac{1}{N_A}\frac{\phi_A^4}{ \phi^4} \right) 
\ea
where $T_{3}$ is the $D$3-brane tension and the effective tension is warped to a very small value $T_{3}h_A^4$.  The warp factor $h_A$ depends on the details of the throat. Crudely, $h(\phi) \sim \phi/\phi_{edge}$, where $\phi=\phi_{edge}$ when the $D$3-brane is at the edge of the throat, so $h(\phi_{edge}) \simeq 1$. At the bottom of the throat, where $\phi=\phi_{A} \ll \phi_{edge}$, $h_{A} = h(\phi_{A})= \phi_{A}/\phi_{edge}$.  The potential is further warped because $N_{A} \gg 1$ is the $D$3 charge of the throat. The $D$3-${\D}$3-brane pair annihilates at $\phi=\phi_{A}$. In terms of the potential (\ref{infpot}), a tachyon appears as $\phi \rightarrow \phi_{A}$ so inflation ends as in hybrid inflation. The energy released by the brane pair annihilation heats up the universe to start the hot big bang. To fit data, $h_{A} \sim 10^{-2}$.  If the last 60 e-folds of inflation takes place inside the throat, then $\phi_{edge} \ge \phi \ge \phi_{A}$ during this period of inflation. 
This simple model yields $n_s \simeq 0.97$, $r < 10^{-5}$ and vanishing non-Gaussianity. 

The above model is very simple and well motivated. There are a number of interesting variations that one may consider. Within the inflaton potential, we can add new features to it.  
In general, we may expect an additional term in $V(\phi)$ of the form 
$$\beta H^2 \phi^{2}/2$$
where $H$ is the Hubble parameter so this interaction term behaves like a conformal coupling.  Such a term can emerge in a number of different ways: \\
$\bullet$ Contributions from the K\"{a}hler potential and various interactions in the superpotential \cite{Kachru:2003sx} as well as possible D-terms \cite{Burgess:2003ic}, so $\beta$ may probe the structure of the flux compactification \cite{Berg:2004sj,Baumann:2006th}.  \\
$\bullet$ It can come from the finite temperature effect. Recall that finite temperature $T$ induces a term of the form $T^2 \phi^2$ in finite temperature field theory. In a de-Sitter universe, there is a Hawking-Gibbons temperature of order $H$ thus inducing a term of the form $H^2\phi^2$.\\
 $\bullet$ The $D$3-brane is attracted to the $\D$3-brane because of the RR charge and the gravitational force. Since the compactified manifold has no boundary, the total RR charge inside must be zero, and so is the gravitational ``charge''. A term of the above form appears if we introduce a smooth background ``charge'' to cancel the gravitational ``charges'' of the branes \cite{buchan_inter-brane_2004}. On the other hand, negative tension is introduced via orientifold planes in the brane world scenario. If the throat is far enough away from the orientifold planes, it is reasonable to ignore this effect.
 
 Overall,  $\beta$ is a free parameter and so is naively expected to be of order unity, $\beta \sim 1$. However, the above potential yields enough inflation only if $\beta$ is small enough, $\beta < 1/5$ \cite{Firouzjahi:2005dh}. The present PLANCK data implies that $\beta$ is essentially zero or even slightly negative.
 
Another variation is to notice that the 6-dimensional throat can be quite non-trivial. In particular, if one treats the throat geometry as a Klebanov-Strassler deformed conifold, gauge-gravity duality leads to the expectation that $\phi$ will encounter steps as the $D$3-brane moves down the throat \cite{Bean:2008na}. Such steps, though small, may be observed in the CMB power spectrum. 

\subsubsection*{Inflection Point Inflation} 

Since the 6-dimensional throat has one radial and 5 angular modes, $V(\phi)$ gets corrections that can have angular dependencies. As a result, its motion towards the bottom of the throat may follow a non-trivial path. One can easily imagine a situation where it will pass over an inflection point. Around the inflection point, $V(\phi)$ may take a generic simple form 
$$V(\phi) \simeq V_0  + A\phi +B\phi^2/2 + C\phi^3/3$$
where $\eta =0$ at the inflection point $B+2C\phi=0$.
As shown in Ref.\cite{Baumann:2006th,Krause:2007jk}, 
given that $V(\phi)$ is flat only around the inflection point, $N_e \sim 1/\sqrt{\epsilon}$, so $\epsilon$ must be very small, resulting in a very small $r$. Here $n_s-1 =2 \eta$ so we expect $n_s$ to be very close to unity, with a slight red or even blue tilt. Although motivated in brane inflation, an inflection point may be encountered in other scenarios of the inflationary universe, so it should not be considered as a stringy feature. 

\subsubsection*{DBI Model}

Instead of modifying the potential, string theory suggests that the kinetic term for a $D$-brane should take the Dirac-Born-Infeld form 
\cite{2004PhRvD..70j3505S,2004PhRvD..70l3505A}
\be
\frac{1}{2}\partial^{\mu}\phi \partial_{\mu}\phi \rightarrow - \frac{1}{f(\phi)}\sqrt{1-f(\phi) \partial^{\mu}\phi \partial_{\mu}\phi} +\frac{1}{f(\phi)}
\ee
where $f(\phi) \sim h(\phi)^{-4} \sim A/\phi^{4}$ is the warp factor of a throat. This DBI property is intrinsically a stringy feature. Here the warp factor plays the role of a brake that slows down the motion of the $D$-brane as it moves down the throat,
$$ f(\phi) (\partial_t\phi)^2 < 1 . $$
This braking mechanism is insensitive to the form of $V(\phi)$, so many e-folds are assured as $\phi \rightarrow 0$.  
 It produces a negligibly small $r$ but a large non-Gaussianity in the equilateral bi-spectrum as the sound speed becomes very small. The CMB data has an upper bound on the non-Gaussianity that clearly rules out the dominance of this stringy DBI effect. Nevertheless, it is a clear example that stringy features of inflationary scenarios may be tested directly by cosmological observation. 
  
 Instead of moving down a throat, one may also consider inflation while a brane is moving out of a throat \cite{Chen:2004gc,Chen:2005ad}. In this case, the predictions are not too different from that of the inflection point case.

 \subsubsection*{$D$3-$D$7-brane Inflation}
 
Here, the inflaton is the position of a $D$3-brane moving relative to the position of a higher dimensional $D$7-brane \cite{Dasgupta:2002ew}. Since the presence of both $D$3 and $D$7-branes preserves supersymmetry (as opposed to the presence of $D$5-branes),  inflation can be driven by a D-term and a potential of the form like 
$$ V(\phi) = V_0 + a \ln \phi - b \phi^2 + c\phi^4 + . . . .$$
 may be generated. Such a potential yields $n_s \simeq 0.98$, which may be a bit too big. One may lower the value a little by considering variations of the scenario. In any case, one ends up with a very small $r$.

\subsection{K\"ahler Moduli Inflation}

In flux compactification, the K\"ahler moduli are typically lighter than the complex structure moduli. Intuitively, this implies their effective potentials are flatter than those for the complex structure moduli. For a Swiss-cheese like compactification, besides the modulus for the overall volume, we can also have moduli describing the sizes of the holes inside the manifold. So it is reasonable to find situations where a K\"ahler modulus plays the role of the inflaton.   A potential is typically generated by non-perturbative effects, so examples may take the form \cite{Conlon:2005jm}
\be
V \simeq V_0 \left(1 - A e^{-k \phi}\right)  
\ee
where both $A$ and $k$ are positive and of order unity or bigger. Here, $\eta \simeq -Ak^2e^{-k\phi}<0$ and $\epsilon \simeq \eta^2/2 k^2$, so the potential is very flat for large enough $\phi$. If the inflaton measures the volume of a blow-up mode corresponding to the size of a 4-cycle in a Swiss-Cheese compactification in the large volume scenario, we have $k \sim \sqrt{\cal V} \ln \cal V$, where the compactification volume is of order ${\cal V} \sim 10^6$ in string units, so $k$ is huge and $r \simeq 2(n_s-1)^2/k^2 \sim 10^{-10}$.
Models of this type are not close to saturating the Lyth bound (\ref{Lyth}).
Other scenarios \cite{Burgess:2013sla} of this type again have small $r$.  It will be interesting whether this type of models can be modified to have a larger $r$.


\pad
\section{Large $r$ Scenarios}

The Lyth bound (\ref{Lyth}) implies that large $r$ requires large inflaton field range $\Delta \phi \gg M_{Pl}$. A phenomenological model with relatively flat potential and large field range is not too hard to write down, e.g. chaotic inflation. However, the range of a typical modulus in string theory is limited by the size of compactification, $\Delta \phi < M_{Pl}$. Even if we could extend the field range (e.g. by considering an irregular shaped manifold), the corrections to a generic potential may grow large as $\phi$ explores a large range. Essentially, we lose control of the approximate description of the potential. Axions allow one to maintain control of the approximation used while exploring large ranges.
Here we shall briefly review 2 ideas, namely the Kim-Nilles-Peloso Mechanism \cite{Kim:2004rp} and the axion monodromy 
\cite{2008PhRvD..78j6003S,McAllister:2008hb}. In both cases, an axion moves in a helical-like path. The flatness of the potential and the large field range appear naturally.



\subsection{The Kim-Nilles-Peloso Mechanism}

Let us start with a simple model and then build up to a model that is relatively satisfactory.

\subsubsection*{Natural Inflation}

It was noticed long ago that axion fields may be ideal inflaton candidates because an axion field $\phi$, a pseudo-scalar mode,  has a shift symmetry,
$\phi \rightarrow \phi \, +$ constant. This symmetry is broken to a discrete symmetry by some non-perturbative effect so that a periodic potential is generated 
\cite{1990PhRvL..65.3233F,1993PhRvD..47..426A},
\be
\label{Natural}
V(\phi) = A \left(1 - \cos \left(\frac{\phi}{f} \right) \right)  \rightarrow \frac{A}{2f^2} \phi^2 + . . . \sim \frac{m^2}{2}\phi^2
\ee
where we have set the minima of the potential at $\phi=0$ to zero vacuum energy. With suitable choice of $A$ and $f$ the resulting axion potential can be relatively small and flat, an important property for inflation. As $f$ becomes large, this model approaches the (quadratic form) chaotic inflation \cite{Linde:1983gd}, which is well studied. To have enough e-folds, we may need a large field range, say, $\Delta \phi > 14 M_{Pl}$, which is possible here only if the decay constant $f > \Delta \phi \gg M_{Pl}$. This requires a certain degree of
fine tuning since a typical $f$ is expected to satisfy $f < M_{Pl}$
(see, for example, Fig. 2 in \cite{Freese:2014nla}).
To fit the scalar mode perturbations of COBE $m \simeq 7 \times 10^{-6}M_{Pl}$. The range of predictions \cite{Freese:2014nla} in the $r$ vs $n_s$ plot is shown in Fig. 2. 

\subsubsection*{N-flation}

\def\assistedinflation{
1998PhRvD..58f1301L,
2001NuPhB.614..101M}
There are ways to get around this to generate enough e-folds. One example is to extend the model to include $N$ different axions
(similar to the idea of using many scalar fields \cite{\assistedinflation})
each with a term of the form in Eq.(\ref{Natural}), with a different decay constant $f_i < M_{Pl}$. This N-flation model \cite{Dimopoulos:2005ac} is a string theory inspired scenario since flux compactification results in the presence of many axions. In the approximation that the axions are independent of each other (that is, their couplings with each other are negligible),  the analysis is quite straightforward and interesting \cite{Easther:2005zr}. The predictions are similar to that of a large field model, that is, $0.93 < n_s < 0.95$ but $r \le 10^{-3}$.
In general, the axions do couple to each other and the situation can be quite complicated. For a particularly simple case,
inspired by string theory and supergravity, a statistical analysis has been carried out and clear predictions can be made \cite{Easther:2005zr}.
For a large number of axions, one has
\be
V(\phi_i)= V_0 + \sum_i \alpha_i \cos\left(\frac{\phi_i}{f_i}\right) + \sum_{i,j} \beta_{ij} \cos\left(\frac{\phi_i}{f_i}-\frac{\phi_j}{f_j}\right) .
\ee
The statistical distribution of results yields small values for $r$ quite similar to that of the N-flation model.

\begin{figure}
\begin{center}
\includegraphics[width=10 cm]{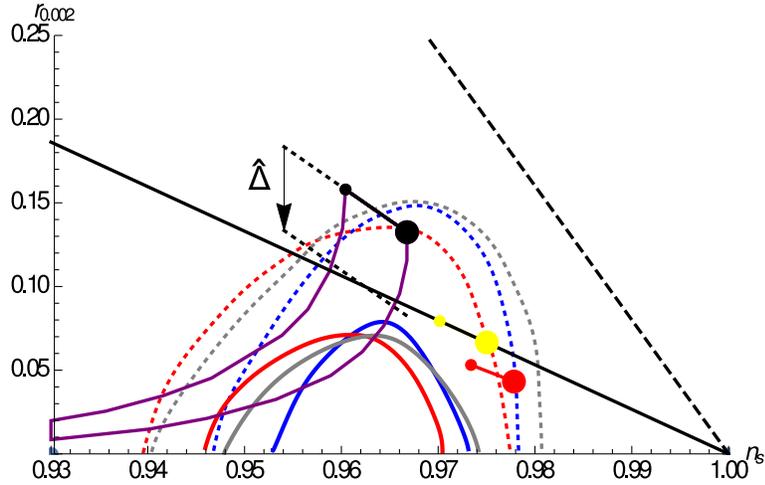}
\vspace{0.1in}
\caption{
The tensor-to-scalar ratio $r$ versus the
primordial tilt $n_s$ plot for pivot scale $0.002$ Mpc$^{-1}$. 
Pairs of solid and dashed semi-circular lines encompass
68\% and 95\% confidence limits for $r$ and $n_s$ as
given in Ref\cite{Ade:2013uln}. 
The regions are based on PLANCK data combined with: WMAP [large scale
  polarization] (grey), WMAP plus BAO [baryon acoustic oscillations in
  galaxy surveys] (blue), and WMAP plus higher $\ell$ CMB data [ACT
  and SPT] (red) as described in detail in Ref\cite{Ade:2013uln}.
All remaining lines, points and arrows
describe theoretical models. Convex potentials
lie above the straight, solid black line; concave potentials below it.
Power law inflation generates results along the straight, dashed line.
String-related models include the short yellow segment 
for a linear inflaton potential $V \sim \phi$,
the short red segment for $V \sim \phi^{2/3}$ and the short black segment
(chaotic inflation) for $V \sim \phi^2$. 
Each small (large) dot stands for 50 (60) e-folds. 
Two purple lines moving to the left show the full range of predictions for natural inflation.
All small field models and some large field ones suppress $r$ (these
are not plotted).
The black arrow is the vertical shift $\hat \Delta$, illustrating
how the results for $cosine$ potential differ from those of
the $\phi^2$ potential. 
}
\label{1.3}
\end{center}
\end{figure}

\subsubsection*{Helical Inflation}

Let us consider a particularly interesting case of a 2-axion model of the form \cite{Kim:2004rp}
\begin{equation}
     V(\phi_1,\phi_2) = V_1+V_2=V_0\left\{ 1- \cos\left({\phi_1\over f_1}\right) +   A\left[1-\cos\left( {\phi_1 \over f_1'}-{\phi_2\over f_2} \right)  \right] \right\} \label{poten}
\end{equation}
where we take $f_1' \ll f_1 \ll M_{Pl}$.


Note that the second term in $V$ (\ref{poten}), namely $V_2$, vanishes at
\begin{equation}
\label{trough}
         {\phi_1 \over f_1'}-{\phi_2\over f_2} = 0 .
\end{equation}
This is the minimum of $V_2$ and the bottom of the trough of the potential $V$.
For large enough $A$, inflaton will follow the trough as it rolls along its path. 
Now $ {f_1 \over 2\pi f_1'}$ measures the number of cycles that $\phi_2$ can travel for $0<\phi_1 < f_1$. Although $\phi_2$ has a shift symmetry $\phi_2 \rightarrow \phi_2 + 2\pi f_2$, the path does not return to the same configuration after $\phi_2$ has traveled for one period because $\phi_1$ has also moved. Thus, instead of a shift symmetry, the system has a helical symmetry.
Moving along the path of the trough (\ref{trough}), we see that the second term in the potential (\ref{poten}) vanishes and so the potential $V$ reduces to that with only the first term and the range of the inflaton field $\phi_2$ can easily be super-Planckian. 

To properly normalize the fields, one defines two normalized orthogonal directions,
\begin{equation}
    X = { f_1' \phi_1 + f_2 \phi_2 \over \sqrt{f_1'^2 + f_2^2}  },   \quad    Y= { f_2 \phi_1 - f_1' \phi_2 \over \sqrt{f_1'^2 + f_2^2}  },
\end{equation}
the inflaton will roll along the $X$-direction while $Y$-direction is a heavy mode which can be integrated out. Note that the system is insensitive to the magnitude of $A$ as long as it is greater than ${\cal O}(1) $ such that $Y$-direction is heavy enough. For instance, the slow roll parameters $\epsilon$ and $\eta$ are not affected which means that the observables $n_s$ and $r$ are insensitive to $A$. 
Since the $X$ path is already at the minimum of the second term, at $Y=0$, the effective potential along $X$ is determined by $V_1$,
\ba
\label{Xcosine}
V(X) & = & V_0\left\{ 1- \cos\left({X\cos \theta /f_1}\right) \right\} \approx  {1\over2}  {V_0 X^2\over f_1^2} \cos^2 \theta\\
 \cos{\theta} & \equiv & {  f_1'  \over \sqrt{f_1'^2 + f_2 ^2}} .
\ea
So this helical model is reduced to the single $cosine$ model (\ref{Natural}) where 
$$f =( f_1/f_1')\sqrt{ f_1'^2 + f_2 ^2} \simeq f_1 f_2/f_1'$$ can be bigger than $M_{Pl}$ even if all the individual $f_i < M_{Pl}$.

So it is not difficult to come up with stringy models that can fit the existing data. In view of the large B-mode reported by BICEP2 
\cite{Ade:2014xna}, 
this model was revisited by a number of groups \cite{Choi:2014rja,Tye:2014tja,Kappl:2014lra,Ben-Dayan:2014zsa,Long:2014dta,Harigaya:2014eta,Higaki:2014pja,Bachlechner:2014hsa}.
In fact, one can consider a more general form for $V_1(\phi_1)$ \cite{Berg:2009tg}, which can lead to a large field range model with somewhat different predictions.

Since the shift symmetry of an axion is typically broken down to some discrete symmetry, the above $cosine$ model is quite natural.
The generalization to more than 2 axions is straightforward, and one can pile additional helical motions on top of this one, increasing the value of the effective decay constant $f$ by additional big factors. All models in this class
reduce to a model with a single $cosine$ potential, which in turn resembles the quadratic version of chaotic inflation. This limiting behavior is quite natural in a large class of axionic models in the supergravity framework \cite{Kallosh:2014vja}. Because of the periodic nature of an axionic potential, this $cosine$ model can have a  smaller value of $r$ than chaotic inflation \cite{Linde:1983gd}. The deviations from chaotic inflation occur in a well-defined fashion. Let
\be
\label{deltah}
{\hat \Delta} = 16 \Delta =  r+ 4(n_s-1) =-4M_{Pl}^2/f^2
\ee
where $\phi^2$ chaotic inflation has $\hat \Delta=0$ while the $cosine$ model has $\hat \Delta <0$. All other quantities such as runnings of spectral indices have very simple dependencies on $\hat \Delta$. As shown in Table 1, this deviation is quite distinctive of the periodic nature of the inflaton potential.
For $n_s=0.96$, $r=0.16$ in the $\phi^2$ model, while $r$ can be as small as $r= 0.04$ (or ${\hat \Delta}= - 0.12$) in the $cosine$ model.
As data improves, a negative value of $\hat \Delta$ can provide a distinctive signature for a periodic axionic potential for inflation.

\begin{center}
   \begin{tabular}{c| c c c c }
        $V(\phi) $  &&                                        ${1\over2} m^2 \phi^2$       &&  $V_0\left[ 1-\cos\left( {\phi \over f} \right) \right] $       \\  \hline   
         $\epsilon =  {1\over2}\left({V' / V}\right)^2 $   &&      ${1\over 16}r$  &&        $   {1\over 16} r $      \\
  $\eta=  {V'' / V}$		&&      ${1\over 16}r$  &&        $   {1\over 16}r + 2\Delta $\\
  $\xi^2 = {V'V''' / V^2}  $  		&&                  0             &&            $ {1\over 2} r\Delta $  \\
  $ \omega^3 = {V'^2V'''' / V^3} $          &&             0             &&       $  {1\over32} r^2\Delta +r\Delta^2  $ \\
  $n_s -1 = 2\eta - 6\epsilon $          &&       $ -{1\over4}r  $            &&                       $  -{1\over4} r + 4\Delta $   \\
  $n_t =-2 \epsilon $ &&   $ -{1\over8}r  $ &&  $ -{1\over8}r  $ \\
$  { dn_s \over d \ln k} = -16\epsilon\eta + 24 \epsilon^2 +2\xi^2  $   &&        ${1\over 32} r^2 $        &&  $ {1\over 32} r^2 - r\Delta   $ \\
$  { dn_t \over d \ln k} = -4\epsilon\eta + 8 \epsilon^2   $   &&      ${1\over 64} r^2 $        &&  $ {1\over 64} r^2 -{1\over2} r\Delta  $ \\
 $ { d^2n_s \over d \ln k^2} = -192\epsilon^3 +192\epsilon^2 \eta -32\epsilon \eta^2 $ &&    $  -{1\over 128 } r^3 $   &&    $  -{1\over 128 } r^3 + {3\over 8}r^2 \Delta -4r \Delta^2 $     \\
 $ -24 \epsilon \xi^2 +2 \eta \xi^2 +2 \omega^3 $   && && \\
   \end{tabular}\\
 \end{center}

 \begin{center}
     {\small{} TABLE 1, Comparison between $\phi^2$ chaotic inflation and the $cosine$ model for the various physically measurable quantities \cite{Tye:2014tja}} where $\Delta$ (\ref{deltah}) measures the difference.
    Here $M_{Pl}=1$.
     \end{center}

\subsection{Axion Monodromy}

The above idea of extending the effective axion range can be carried out with a single axion, where the axion itself executes a helical motion. This is the axion monodromy model 
\cite{2008PhRvD..78j6003S,McAllister:2008hb}. In this scenario, inflation can persist through many periods around the configuration space, thus generating an effectively large field range with an observable $r$.

Recall that  a gauge field, or one-form field (i.e., with one space-time index), is sourced by charged point-like fields, while a 2-form (anti-symmetric tensor) field is sourced by strings. So we see that there will be at least one 2-form field in string theory. Consider a 5-brane that fills our 4-dimensional space-time and wraps a 2-cycle inside the compactified manifold. 
The axion is the integral of a 2-form field over the 2-cycle. Integrating over this 2-cycle, the 6-dimensional brane action produces a potential for the axion field in the resulting 4-dimensional effective theory. Here, the presence of the brane breaks the axion shift symmetry and generates a monodromy for the axion.  For NS5-branes, a typical form of the axion potential (coming from the DBI action) is
\be
V(\phi) = A \sqrt{b^2 + \phi^2} .
\ee
For small parameter $b$, $V(\phi) \simeq A\phi$. The prediction of such a linear potential is shown in Fig. 2.
One can consider $D$5-branes instead.  $D$7-branes wrapping 4-cycles in the compactified manifold is another possibility.
 For large values of $\phi$, a good axion monodromy model requires that there is no uncontrollable higher order stringy or quantum corrections that would spoil the above interesting properties
\cite{2014arXiv1404.3040M,McAllister:2014mpa,Arends:2014qca}.
Variants of this picture may allow a more general form of the potential, say $V(\phi) \sim \phi^p$ where $p$ can take values such as $p=2/3, 4/3, 2, 3$, thus generating $r \simeq 0.04,0.09, 0.13$ and $0.2$, respectively.
The $p=2$ case predicts a $n_s$ value closest to $r \sim 0.16$ reported
by BICEP2 after accounting for dust. The observational bounds are
currently under scrutiny.
Readers are referred to Ref\cite{Baumann:2014nda} for more details.

\subsection{Discussions}

If BICEP2's detection of $r$ is confirmed, it does not necessarily invalidate completely all the small $r$ models discussed above. It may be possible to modify some of them to generate a large enough $r$. As an example, if one is willing to embed the $D$3-${\D}$3-brane inflation into a warm inflationary model, a $r \simeq 0.2$ may be viable \cite{Setare:2014qea}. It will be interesting to re-examine all the small $r$ string theory models and see whether and how any of them may be modified to produce a large $r$.

As string theory has numerous solutions, it is not surprising that there are multiple ways to realize the inflationary universe scenario. With cosmological data available today, theoretical predictions and contact with observations are so far  quite limited. As a consequence, it is rather difficult to distinguish many string theory inspired predictions from those coming from ordinary field theory (or even supergravity models). There are exceptions, as pointed out earlier. For example, the DBI inflation prediction of an equilateral bi-spectrum in the non-Gaussianity, or the determined spacing of steps in the power spectrum itself, may be considered to be distinct enough that if either one is observed, some of us may be convinced that it is a smoking gun of string theory.  Although searches for these phenomena should and would continue, so far, we have not been lucky enough to see any hint of them.

Here we like to emphasize that there is another plausible signature to search for. Since all fundamental objects are made of superstrings (we include $D$1-strings here), and the universe is reheated to produce a hot big bang after inflation, it is likely that, besides strings in their lowest modes which appear as ordinary particles,  
some relatively long strings will also be produced, either via the Kibble mechanism or some other mechanism. They will appear as cosmic strings.

When there are many axions, or axion-like fields, with a variety of plausible potentials, the possibilities may be quite numerous and so predictions may be somewhat imprecise. For any axion or would-be-Goldstone bosons, with a continuous $U(1)$ symmetry, we expect a string-like defect which can end up as cosmic strings if generated in early universe.  In field theory, a vortex simply follows from the Higgs mechanism where the axion $a$ appears as the phase of a complex scalar field, $\Phi=\rho e^{i a/f}$.  In string theory, we note that such an axion $a$ is dual to a 2-form tensor field $C_{\lambda \kappa}$, i.e., $\partial^{\mu}a =\epsilon^{\mu \nu \lambda \kappa} \partial_{\nu} C_{\lambda \kappa}$.
 As pointed out earlier, such a 2-form field is sourced by a string. So the presence of axions would easily lead to cosmic strings (i.e., vortices, fundamental strings and D1-strings) and these may provide signatures of string theory scenarios for the inflationary universe. This is especially relevant if cosmic strings come in a variety of types with different tensions and maybe even with junctions.

 Following Fig. 1, we see that besides the axions responsible for inflation, there may be other axions. Some of them may have mass scales warped to very small values.  
If a potential of the form (\ref{Natural}) is generated, a closed string loop becomes the boundary of a domain wall, or membrane. It will be interesting to study the effect of the membrane on the evolution of a cosmic string loop.
The tension of the membrane and the axion mass are of order 
$$ \sigma \sim \sqrt{A} f,  \quad \quad m^2=A/f^2 $$ 
It is interesting to entertain the possibility that this axion can contribute substantially to the dark matter of the universe. If so, its contribution to the energy density is roughly given by $A$ while its mass is estimated to be $m\simeq 10^{-22}$ eV 
\cite{2014NatPh..10..496S}.
This yields $A \simeq 10^{-118} M_{Pl}^4$, $f \simeq \sqrt{A}/m \simeq  10^{-10} M_{Pl}$ and $\sigma \simeq 10^{-69} M^3_{Pl} \simeq 10^{-14}$ GeV$^3$. For such a small membrane tension, the evolution of the corresponding cosmic string is probably not much changed.

\pad  
\section{Relics: Low Tension Cosmic Strings}

Witten's \cite{witten_cosmic_1985} original consideration of
macroscopic cosmic strings was highly influential. He argued that the
fundamental strings in heterotic string theory had tensions too large
to be consistent with the isotropy of the COBE observations. Had they
been produced they would be inflated away. It is not even clear how
inflation might be realized within the heterotic string theory.

However, with the discovery of $D$-branes
\cite{polchinski_dirichlet_1995}, the introduction of warped
geometries \cite{randall_large_1999} and the development of specific,
string-based inflationary scenarios, the picture has changed
substantially. Open fundamental strings must end on branes; so both
open and closed strings are present in the brane world. Closed string
loops inside a brane will break up into pieces of open strings, so
only vortices (which may be only meta-stable) can survive inside
branes. Here, $D$1-strings (i.e., $D$1-branes) may be treated as vortices inside branes
and survive long enough to be cosmologically interesting \cite{Leblond:2004uc}. 

The warped geometry will gravitationally redshift the string tensions
to low values and, consequently, strings can be produced {\it after}
inflation. A string with tension $T$ in the bulk will be warped to 
$\mu =h^2T$ with $h$ being the warp factor, which can be very small when the string is sitting at the bottom of a throat. (It is $h=h_A$ in Eq.(\ref{infpot}) in throat A.)  
That is important because relics produced during inflation
are rapidly diluted by expansion. Only those generated after (or
very near the end of) inflation are potentially found within the
visible universe.  Since the Type IIB model has neither $D$0-branes nor $D$2-branes, the well-justified conclusion that our universe is dominated
neither by monopoles nor by domain walls follows automatically. 
Scenarios that incorporate string-like relics may prove to be
consistent with all observations. Such relics appear to be natural
outcomes of today's best understood string theory scenarios.

The physical details of the strings in Type IIB model can be quite non-trivial,
including the types of different species present and the range of
string tensions.  Away from the branes, $p$ fundamental F1-strings and
$q$ $D$1-strings can form a $(p,q)$ bound state.
Junctions of strings will be present automatically. If they live at the bottom of a throat, there
can be beads at the junctions as well.
To be specific, let us consider the case of the Klebanov-Strassler warped throat \cite{Klebanov:2000hb},  whose properties are relatively well understood. On the gravity side, this is a warped deformed conifold. Inside the throat, the geometry is a shrinking $S^{2}$ fibered over a $S^{3}$.
The tensions of the bound state of $p$ F1-strings and that of $q$ $D$1-strings were individually computed \cite{Gubser:2004qj}. 
The tension formula for the $(p,q)$ bound states is given by \cite{Firouzjahi:2006vp} in terms of the warp factor $h$, the strings scale $M_S$ and the string coupling $g_s$,
\be
\label{finaltension}
T_{p,q} \simeq  \frac{h^{2}M_S^2}{2 \pi} \sqrt{\frac{q^2}{g_s^2} + \left(\frac{b M}{\pi}\right)^2 \sin^2\left(\frac{\pi p}{M}\right)},
\ee  
where $b=0.93$ numerically and $M$ is the number of fractional D3-branes (that is, the units of 3-form RR flux $F_3$ through the $S^{3}$).
Interestingly, the $F$1-strings are charged in $\mathbb{Z}_M$ and are non-BPS. 
The D-string on the other hand is charged in $\mathbb{Z}$ and is BPS with respect to each other.  
Because $p$ is $Z_{M}$-charged with non-zero binding energy, binding can take place even if $(p,q)$ are not  coprime. Since it is a convex function, i.e., $T_{p+p'} < T_{p} + T_{p'}$, the $p$-string will not decay into strings with smaller $p$. $M$ fundamental strings can terminate to a point-like bead with mass \cite{Gubser:2004qj}
$$M_{bead} = \sqrt{\frac{g_s}{4\pi}}\left(\frac{bM}{\pi}\right)^{3/2}  \frac{h}{3}M_S$$ irrespective of the number of D-strings around. Inside a $D$-brane, F1-strings break into pieces of open strings so we are left with $D$1-strings only, in which case they resemble the usual vortices in field theory. For $M \rightarrow \infty$ and $b=h=1$, the tension (\ref{finaltension}) reduces to that for flat internal space \cite{copeland_cosmic_2004}. Cosmological properties of the beads have also been studied \cite{2007PhRvD..75l3522L}.

Following from gauge-gravity duality, the interpretation of these strings in the gauge theory dual is known. The $F$1-string is dual to a confining string  while the $D$1-string is dual to an axionic string.  Here the gauge theory is strongly interacting and the bead plays the role of a ``baryon".  
It is likely that different throats have different tension spectra similar to that for the Klebanov-Strassler throat.
Finally, there is some evidence that strings can move in both internal
and external dimensions and are not necessarily confined to the tips
of the throat \cite{avgoustidis_cosmic_2012}. This behavior can show
up as a cosmic string with variable tension. Following gauge-gravity duality, one may argue 
that one can also obtain these types of strings within a strongly interacting gauge theory; 
however, so far we are unable to see how inflation can emerge from such a 
non-perturbative gauge theory description.

As we will describe in more detail, all indications suggest that, once
produced after inflation, a scaling cosmic string network will emerge
for the stable strings.
Before string theory's application to cosmology, the typical
cosmic string tension was presumed to be set by the grand
unified theory's (GUTs) energy scale. Such strings have been
ruled out by observations. We will review the current
limits shortly. Warped geometry typically allows strings with 
tensions that are substantially smaller, avoiding the
observational constraints on the one hand and frustrating
easy detection on the other. The universe's expansion
inevitably generates sub-horizon string loops and if the
tension is small enough the loops will survive so long that
their peculiar motions are damped and they will cluster in
the manner of cold dark matter. This results in a huge
enhancement of cosmic string loops within our galaxy (about 
$10^5$ times larger at the Sun's position 
than the mean throughout the universe) and
makes detection of the local population a realistic
experimental goal in the near future. We will focus on the
path to detection by means of microlensing. Elsewhere, we will
discuss blind, gravitational wave searches.

A microlensing detection will be very distinctive. The nature of a
microlensing loop can be further confirmed and studied through its
unique gravitational wave signature involving emission of multiple
harmonics of the fundamental loop period from the precise microlensing
direction. Since these loops are essentially the same type of string
that makes up all forms of microscopic matter in the universe, their
detection will be of fundamental importance in our understanding of
nature.

\subsection{Strings in Brane World Cosmology}

String-like defects or fundamental strings are expected
  whenever reheating produces some
  closed strings towards the end of an inflationary
  epoch.
Once inflation ends and the radiation dominated epoch begins,
ever larger sections of this cosmic string network re-enter
the horizon. The strings move at relativistic speeds and
long lengths collide and break off sub-horizon loops. Loops
shrink and evaporate by emitting gravitational waves in a
characteristic time $\tau = \linvariant/(\Gamma G \mu)$ where $\linvariant$ is
the invariant loop size, $\mu$ is the string tension while $\Gamma$ is numerically determined and $\Gamma \sim 50$ for strings coupled only to gravity \cite{vilenkin_cosmic_2000}. To ease discussion, we shall adopt this value for $\Gamma$.

String tension is the primary parameter that controls the cosmic string
network evolution, first explored in the context
of phase transitions in grand unified field theories
(GUTs), which may be tied to the string scale. Assuming that the inflation scale is comparable to the GUT scale, inflation generated horizon-crossing defects
whose tension is
set by the characteristic grand unification energy
\cite{vilenkin_cosmic_2000}.
These GUT strings with $G \mu \sim
10^{-6}$ would have seeded the density fluctuations for galaxies
and clusters but have long been ruled out by
observations of the cosmic microwave background (CMB)
\cite{smoot_structure_1992,bennett_four-year_1996,spergel_three-year_2007}.

Here, string theory comes to the rescue. Six of the string
theory's 9 spatial dimensions are stably compactified. The flux
compactification involves manifolds possessing warped
throat-like structures which redshift all characteristic
energy scales compared to those in the bulk space. In this
context, cosmic strings produced after inflation living in or near the bottoms of the throats can have different small tensions  
\cite{jones_brane_2002,sarangi_cosmic_2002-1,jones_production_2003,firouzjahi_brane_2005,shandera_observing_2006}.
The quantum theory of one-dimensional
objects includes a host of {\it effectively} one-dimensional
objects collectively referred to here as superstrings. 
For example, a single $D$3-brane has a $U(1)$ symmetry that is expected to be broken, thus generating a string-like defect. (For a stack of $n$ branes, the $U(n) \supset U(1)$ symmetry is generic.)
Note that fundamental superstring loops can exist only away from branes. Any superstring we observe will have tension $\mu$
exponentially diminished from that of the Planck
scale by virtue of its location at the bottom
of the throat. Values like $G \mu < 10^{-14}$ (i.e. energies $<
10^{12}$ GeV) are entirely possible. A typical
manifold will have many throats and we expect a distribution
of $\mu$, presumably with some $G \mu >
10^{-14}$. 

In addition to their reduced tensions, superstrings should differ
from standard field theory strings (i.e., vortices) in other important ways:
long-lived excited states with junctions and beads may
exist, multiple non-interacting species of superstrings may
coexist and, finally, the probability for breaking and
rejoining colliding segments (intercommutation) can be much
smaller than unity \cite{jackson_collisions_2005}. Furthermore a closed string loop may move inside the compacted volume as well. Because of the warped geometry there, such motion may be observed as a variable string tension both along the string length and in time.

\subsection{Current Bounds on String Tension $G\mu$ and Probability of Intercommutation $p$}
\def\lensinglimits{
vilenkin_gravitational_1983,
hogan_gravitational_1984-1,
vilenkin_cosmic_1984,
de_laix_observing_1997,
bernardeau_cosmic_2001,
sazhin_csl-1:_2003,
sazhin_true_2006,
2008PhRvD..77l3509C,
christiansen_search_2011}
\def\gravitationalwavebackandburst{
hogan_gravitational_1984-2,
vachaspati_gravitational_1985,
economou_gravitational_1992,
battye_gravitational_1998,
damour_gravitational_2000,
damour_gravitational_2001,
damour_gravitational_2005,
siemens_gravitational_2006-1,
hogan_gravitational_2006,
siemens_gravitational-wave_2007,
abbott_searches_2007-short,
ligo_scientific_collaboration:_b._abbott_first_2009,
abbott_upper_2009-short,
aasi_constraints_2014}
\def\gravitationalwavepulsars{
bouchet_millisecond-pulsar_1990,
hogan_gravitational_1984-1,
caldwell_cosmological_1992,
kaspi_high-precision_1994,
jenet_upper_2006,
depies_stochastic_2007,
demorest_limits_2013}
\def\tensionlimits{
smoot_structure_1992,
bennett_four-year_1996,
pogosian_observational_2003,
pogosian_observational_2004,
tye_scaling_2005,
wyman_bounds_2005,
pogosian_vector_2006,
seljak_cosmological_2006,
spergel_three-year_2007,
bevis_cmb_2007,
fraisse_limits_2007,
bevis_cosmic_2008,
pogosian_cosmic_2009,
battye_updated_2010,
dunkley_atacama_2011-short,
Ade:2013uln,
2013arXiv1303.5085P}

Empirical upper bounds on $G \mu$ have been derived from null results
for experiments involving lensing \cite{\lensinglimits}, gravitational
wave background and bursts \cite{\gravitationalwavebackandburst},
pulsar timing \cite{\gravitationalwavepulsars} and cosmic microwave
background radiation \cite{\tensionlimits}. We will briefly review
some recent results but see 
\cite{copeland_seeking_2011,sanidas_constraints_2012} for more
comprehensive treatments.

All bounds on string tension depend upon uncertain aspects of string physics
and of network modeling. The most important factors include:
\begin{itemize}
\item The range of loop sizes generated by network evolution. ``Large'' means
comparable to the Hubble scale, ``small'' can be as small as the core width of the
string. Large loops take longer to evaporate by emission of gravitational
radiation. 
\item The probability of intercommutation $p$ is the probability that
two crossing strings break and reconnect to form new continuous segments. Field theory strings
have $p \sim 1$ but superstrings may have $p$ as small as $10^{-3}$, depending on the crossing angle and the relative speed.
The effect of lowering $p$ is to increase
the network density of strings to maintain scaling.
\item The physical structure of the strings. F1 strings are one-dimensional,
obeying Nambu-Goto equations of motion. $D$1 strings and field theory strings
are vortices with finite cores.
\item The mathematical description of string dynamics used in 
simulations and calculations. The
Abelian Higgs model is the simplest vortex description but
the core size in calculations is not set to realistic physical values.
Abelian Higgs and Nambu-Goto descriptions yield different
string dynamics on small scales.
\item The number of stable string species. Superstring
have more possibilities than simple field theory strings. These
include bound states of F1
and $D$1 strings with beads at the junctions and non-interacting strings from different warped throats.
\item The charges and/or fluxes carried by the string. Many effectively
one-dimensional objects in string theory can experience
non-gravitational interactions.
\item The character of the discontinuities on a typical loop. The number of
cusps and/or the number of kinks governs the emitted gravitational
wave spectrum. 
\end{itemize}
While there has been tremendous progress, all of these areas
are under active study.

When the model-related theoretical factors are fixed each astrophysical
experiment probes a subset of the string content of spacetime. For example,
CMB power spectrum fits rely on
well-established gross properties of large-scale string networks which
are relatively secure but do not probe small sized loops which
may dominate the total energy density and which would show up only at large
$\ell$. Analysis of combined PLANCK, WMAP, SPT and ACT
data \cite{2013arXiv1303.5085P} implies $G \mu < 1.3 \times 10^{-7}$
for Nambu-Goto strings and $G \mu <3.0 \times 10^{-7}$ for field theory
strings. Limits from optical lensing in fields of background galaxies
rely on the theoretically well-understood deficit angle geometry of a
string in spacetime but require a precise accounting for observational
selection effects. Analysis of the GOODS and COSMOS optical
surveys \cite{2008PhRvD..77l3509C,christiansen_search_2011} yields
$G \mu < 3 \times 10^{-7}$.  Taken together these observations
imply $G \mu \lta 1 \text{--} 3 \times 10^{-7}$.

There is a well-established bound on the gravitational energy density
at the time of Big Bang Nucleosynthesis because an altered expansion
rate impacts light element yields (e.g. \cite{buonanno_tasi_2003}). The
gravitational radiation generated by any string network cannot exceed the
bounds. If a network forms large Nambu-Goto loops of one type of
string with intercommutation
probability $p$ an estimate of the BBN constraint is $G\mu \lta 5 \times
10^{-7} p^2$ \cite{oelmez_gravitational-wave_2010} (this depends
implicitly on the loop formation size and a still-emerging
understanding of how network densities vary with $p$). 
LIGO's experimental bound on
the stochastic background gravitational radiation \cite{abbott_upper_2009-short}
from a similar network
implies a limit on $G\mu$ of the same general form as the BBN limit but
weaker. Advanced LIGO is projected to reach $G\mu \sim 10^{-12}$ for $p=1$
\cite{oelmez_gravitational-wave_2010}. The same LIGO results can be used
to place a reliable, conservative bound of
$G\mu < 2.6 \times 10^{-4}$ over a much wider range of
possible models, almost independent of loop size and of
frequency scaling of the emission \cite{sanidas_constraints_2012}.

\begin{figure}
\begin{center}
\includegraphics[width=9cm]{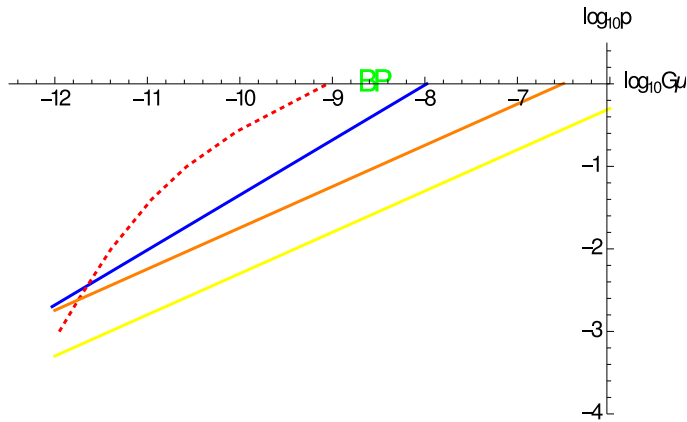}
\vspace{0.1in}
\caption{Observational constraints limit
string tension and intercommutation
probability if the network forms large loops. Each line is
a particular observational constraint. The
region below and to the right of a line is disfavored. Smaller
$p$ increases the number density of loops and larger $G \mu$
increases gravitational signal amplitude. Conversely, the region to the left
and above a line remains consistent with observational limits.
The illustrated constraints are:
Big Bang Nucleosynthesis (orange line) and
CMB (yellow line) from the gravitational wave
spectra in \cite{oelmez_gravitational-wave_2010},
pulsar time of arrival for the Parkes Pulsar Timing Array (blue line,
based on the analytic form in \cite{jenet_upper_2006}) and
for NANOGrav (red dotted line) in \cite{demorest_limits_2013}.
All lines depend upon theoretical modeling of the network, especially
the fraction of large loops formed. For
example, the initials BP (green) \cite{blanco-pillado_number_2014} are
an example in which this fraction has been inferred from 
simulations and the bound on $G \mu$ at $p=1$ indicated (the line
has not been calculated). This result
is approximately $\sim 30$ times {\it less} restrictive on $G \mu$ than
an equivalent analysis in which all loops are born at a single
large size as has been assumed in all the other analyses.
}
\label{astrophysical-limits}
\end{center}
\end{figure}

More stringent bounds rely on additional assumptions. Consider a specific
set of choices for the secondary parameters of strings: one
string species, Nambu-Goto dynamics, only gravitational
interactions, large loops ($\alpha = 0.1$), each with
a cusp. The time of arrival of pulses emitted by a pulsar vary
on account of the gravitational wave background. When the sequence is
observed to be regular the perturbing amplitude of the waves
is limited. The current limit is $G \mu \lta 10^{-9}$ for
$p=1$ and $\lta 10^{-12}$ for $p=10^{-3}$
\cite{demorest_limits_2013}. Figure \ref{astrophysical-limits}
is a graphical summary that illustrates some of the bounds
discussed.  The BBN
(orange line) and CMB (yellow line) constraints are shown as a
function of string tension $G \mu$ and intercommutation probability
$p$.  The Parkes Pulsar
Timing Array limit (blue line)
\cite{jenet_upper_2006} and the NANOGrav limit (red dotted
line) \cite{demorest_limits_2013} are based on radiating
cusp models. Each constraint rules out the area below and to the right 
of a line. Most
analyses do not account for the fact that only a small fraction
of horizon-crossing string actually form large loops which
are ultimately responsible for variation in arrival times. In this
sense the lines may be over-optimistic (see figure caption).

Superconducting strings have also been proposed \cite{witten1985a}.
The bound on superconducting cosmic strings is about $G \mu \lta 10^{-10}$ 
\cite{miyamoto_cosmological_2013}. Interestingly, it was pointed out that the recently observed fast radio bursts can be consistent with being produced by superconducting cosmic strings \cite{Yu2014}.


In short, cosmic superstrings are generically produced towards the end of
inflation and observations imply tensions substantially less than the
original GUT-inspired strings. 
Multiple, overlapping approaches are needed to minimize physical uncertainties
and model-dependent aspects.
There is no known theoretical impediment to the magnitude of
$G \mu$ being either comparable to or much lower than the current
observational upper limits.

\pad
\section{Scaling, Slowing, Clustering and Evaporating}

The important physical processes are network scaling and
loop slowing, clustering and evaporating.\footnote{Material in this section \cite{chernoff_localmodel_2014}.}
Simulated cosmological string networks converge to
self-similar scaling solutions
\cite{vilenkin_cosmic_2000,jackson_collisions_2005,tye_scaling_2005}.
Consequently $\Omega_{long}$ ($\Omega_{loop}$), the fraction
of the critical density contributed by horizon-crossing
strings (loops), is independent of time while the
characteristic size of a loop formed at time $t$ scales with
the size of the horizon: $\linvariant = \alpha t$ for some fixed
$\alpha$.  To achieve scaling, long strings that enter the
horizon must be chopped into loops sufficiently rapidly --
if not, the density of long strings increases and over-closes
the universe. In addition, loops must be removed so that
$\Omega_{loop}$ stabilizes -- if not, loops
would come to dominate the contribution of normal matter.
Scaling of the string network
is an attractor solution in many well-studied models and
the universe escapes the jaws of both Scylla and Charybdis.
The intercommutation probability determines the efficiency
of chopping and $\mu$ determines the rate of loop
evaporation.

Studies of GUT strings \cite{vilenkin_cosmic_2000} took $G \mu \sim
10^{-6}$ (large enough to generate perturbations of
interest at matter-radiation equality) and $\alpha$ small (set by
early estimates of gravitational wave damping on the long
strings) with the consequence $H \tau \sim
\alpha/\Gamma G \mu << 1$ where $H$ is the Hubble
constant. Newly formed GUT loops decay quickly.
The superstrings of interest here have smaller $G \mu$ so that loops of a
given size live longer. In addition, recent simulations
\cite{vanchurin_cosmic_2005,sakellariadou_note_2005,martins_fractal_2006,
avgoustidis_effect_2006,ringeval_cosmological_2007}
produce a range of large loops: $10^{-4} \lta
\alpha \lta 0.25$.  The best current understanding
is that $\sim 10-20$\% of the long string length that
is cut up goes
into loops comparable to the scale of the horizon
($\alpha \sim 0.1$) while the remaining $\sim 80-90$\%
fragments to much smaller size scales
\cite{polchinski_analytic_2006,polchinski_cosmic_2007-1}.
The newly formed, large loops are the most important
contribution for determining today's loop population.

Cosmic expansion strongly damps the initial
relativistic center of mass motions of the loops and
promotes clustering of the loops as matter
perturbations grow
\cite{chernoff_cosmic_2007,chernoff_clustering_2009}. Clustering was
irrelevant for the GUT-inspired loops. They moved
rapidly at birth, damped briefly by cosmic expansion
and were re-accelerated to mildly relativistic
velocities by the momentum recoil of anisotropic
gravitational wave emission (the rocket effect) before
fully evaporating
\cite{hogan_runaway_1987,hogan_gravitational_1984-3,vachaspati_gravitational_1985}.
GUT loops were homogeneously distributed throughout
space. By contrast, below a critical tension $G \mu
\sim 10^{-9}$ all superstring loops accrete along with
the cold dark matter \cite{chernoff_clustering_2009}.


Loops of size $\linvariant = \lgravcutoff \equiv
\Gamma G \mu t_0$ are just now evaporating where $t_0$ is
the age of the universe. The mean number density of
such loops is dominated by network fragmentation
when the universe was most dense, i.e. at early
times. When they were born they came from the large end of the size
spectrum, i.e.  a substantial fraction of the scale of
the horizon.  The epoch of birth is $t_i = \lgravcutoff
/\alpha =
\Gamma G \mu t_0/\alpha$. For $G \mu < 7 \times 10^{-9}
(\alpha/0.1)(50/\Gamma)$ loops are born before
equipartition in $\Lambda$CDM, i.e.  $t_i <
t_{eq}$. The smallest loops today have $\lgravcutoff \approx 40
{\rm pc} (G \mu/2 \times 10^{-10})$ with characteristic
mass scale $\mgravcutoff = 1.7 \times 10^5 \msun \left( G \mu/2
\times 10^{-10} \right)^2$ for $\Gamma=50$.

Loop number and energy densities today are dominated
by the scale of the gravitational cutoff, the smallest loops that
have not yet evaporated. The characteristic number
density $\dndlogl \propto \left( \Gamma G \mu
\right)^{-3/2} \left( \alpha t_{eq}/t_0 \right)^{1/2}
/t_0^3$ and the energy density $d\rho_{loop}/d \log \linvariant
\sim \Gamma G \mu t_0 \dndlogl$.  The latter implies
$\Omega_{loops} \propto \sqrt{\alpha G \mu/\Gamma}$
whereas $\Omega_{long} \propto \Gamma G \mu$. In scenarios with small
$\mu$ it's the loops that dominate long, horizon-crossing strings
in various observable contexts.  The probability and
rate of local lensing are proportional to the energy
density.

\begin{figure}
\includegraphics[height=3in]{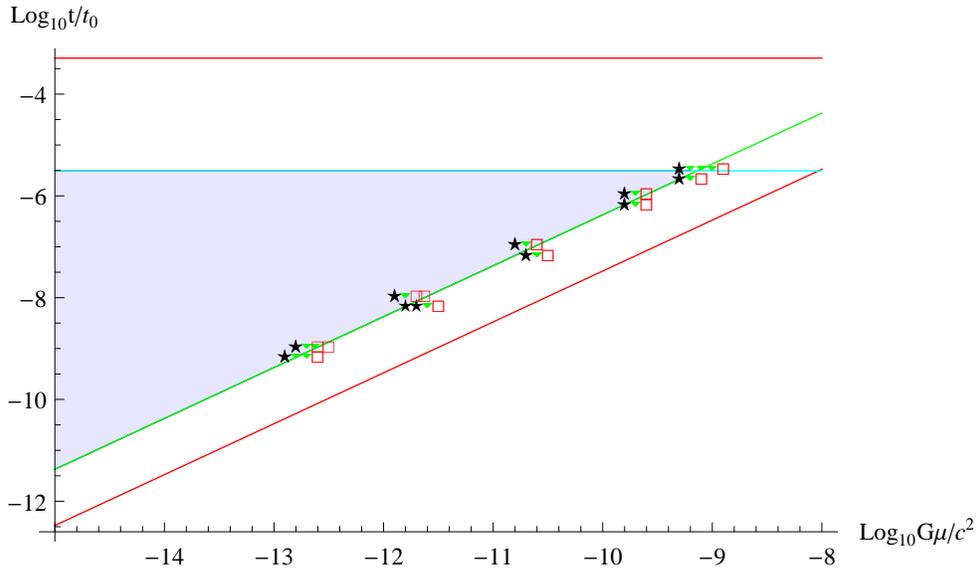}
\caption{\label{fig-summary-ti-mu} Bounds on formation time
and string tension for a loop with initial velocity $v_i=0.1$
to be captured at physical radius $30$ kpc by galaxy formation.
{\bf (1)} Upper bounds on the formation time $t_i/t_0$ are given by the
horizontal lines. The condition
that cosmic drag lower the velocity to less than the circular rotation
velocity today is given by the red line. The more stringent condition that
capture occur at $30$ kpc is given by the turquoise line.
{\bf (2)} Upper bounds on the string tension $G \mu$ are given by the
diagonal lines. The condition that the loop be younger than its
gravitational wave decay timescale is given by the red line.  The more
stringent condition that the loop not be accelerated out of the galaxy
by today is given by the green line. 
{\bf (3)} The shaded region encompasses string tensions and formation times
giving bound loops at $30$ kpc for $v_i=0.1$ and $\alpha=0.1$.
The critical value of $G \mu$ below
which clustering is possible is determined by the upper right hand corner
of the green and turquoise lines.
Lowering $v_i$ raises the limit on
$t_i/t_0$ (horizontal lines moves upward); lowering $\alpha$ shifts the
bound to smaller $G \mu$ (diagonal lines moves leftward). Shifting the
loop orbital scale to smaller values (say the solar position)
requires earlier formation times (horizontal lines shifts down) and
allows larger $G \mu$ (green line shifts to the right but is limited by
the red line which is fixed).
{\bf (4)} The geometric symbols illustrate the sensitivity of the rocket
effect.
These are numerical experiments examining the outcome today ($t=t_0$) for
groups of 10-20 loops captured at $30$ kpc with slightly different
string tensions: stars = all loops bound, boxes = all loops ejected,
triangles = some bound and some ejected (for clarity the points are slightly
offset in the vertical but not the horizontal direction).
}
\end{figure}

Loops are accreted as the galaxy forms. 
Figure \ref{fig-summary-ti-mu} illustrates
schematically the constraints for a loop with typical
initial peculiar velocity to be captured during galaxy
formation and to remain bound today.  The loop must lie
within the inner triangular region which delimits small
enough tension and early enough time of formation.
Above the horizontal line capture is impossible; below
the diagonal line detachment by the rocket effect has
already occurred. The critical tension for loop
clustering in the galaxy is set by the right hand
corner of the allowed region.

To briefly summarize:
${\cal F}$, the enhancement of the galactic
loop number density over the homogeneous mean, simply traces
${\cal E}$, the enhancement of cold dark matter over $\Omega_{DM} \rho_c $,
for critical density $\rho_c$. This
encapsulates conclusions of a study of the growth of a
galactic matter perturbation and the simultaneous capture
and escape of network-generated loops
\cite{chernoff_clustering_2009}. At a typical
galactocentric distance of $11$ kpc $\log_{10} {\bar {\cal
    F}} = \log_{10} {\bar {\cal E}} + f(y)$ where $\log_{10}
{\bar {\cal E}} = 5.5$ and $f(y) = -0.337 - 0.064 y^2 - g(y)$
and $g(y) = 5(1+\tanh(y-7))$. The
tension-dependent deviation from loops as
passive tracers of cold dark matter is $f(y)$, $y=\log_{10}
\left( G \mu/10^{-15} \right)$, fit for $0 \le y \le 5$ from the capture study. 
Clustering saturates for small tensions: $f(y)=f(0)$
for $y<0$.  The over-bar indicates quantities averaged over
the spherical volume and, in the case of the loops, a
weighting by loop length in the capture study.

The extra piece $g(y)$ plays a role for $G \mu > 10^{-10}$ where it
describes the suppression in clustering as one approaches the upper
right corner of the triangle in figure
\ref{fig-summary-ti-mu}. Numerical simulations have not yet accurately
determined it.

A single $f(y)$ fits a range of radii as long as ${\bar
  {\cal E}} >> 1$ (typically, galactic distances less than
$\sim 100$ kpc). So we can easily have an enhancement ${\cal
  F} \sim 10^5$ for $G \mu \sim 10^{-14}$ at the Sun's
position. By comparison, GUT loops would have ${\cal F}=1$
for all positions and tensions.


The model accounts for the local population of strings and
we have used it to estimate microlensing rates, and LISA-like, LIGO-like
and NANOGrav-like burst rates.

\subsection{Large-scale String Distribution}

We will start with a ``baseline'' description (loops from a network of
a single, gravitationally interacting, Nambu-Goto string species with
reconnection probability $p=1$).  The detailed description
\cite{chernoff_clustering_2009} was motivated by analytic arguments
\cite{polchinski_analytic_2006,polchinski_cosmic_2007-1} that roughly
80\% of the network invariant length was chopped into strings with
very small loop size ($\alpha \sim G\mu$) and by the numerical result
\cite{vanchurin_cosmic_2005,sakellariadou_note_2005,martins_fractal_2006,
  avgoustidis_effect_2006,ringeval_cosmological_2007} that the
remaining 20\% formed large, long-lived loops ($\alpha =0.1$).  At a
given epoch loops are created with a range of sizes but only the
``large'' ones are of interest for the local population. The
baseline description is supposed to be directly comparable to
numerical simulations which generally take $p=1$. The most
recent simulations \cite{blanco-pillado_number_2014} are qualitatively
consistent.

Next, we parameterize the actual ``homogeneous'' distribution in the
universe when string theory introduces a multiplicity of string
species and the reconnection probability $p$ may be
less than 1.  And finally we will form the ``local'' distribution which
accounts for the clustering of the homogeneous distribution.

In a physical volume $V$ with a network of long,
horizon-crossing strings of tension $\mu$ with persistence
length $L$ there are $V/L^3$ segments of length $L$. The
physical energy density is $\rho_\infty = \mu L/V =
\mu/L^2$.  The persistence length evolves as the universe
expands. A scaling solution demands $L \propto t$ during
power law phases.  There are also loops within the horizon;
their energy is not included in $\rho_\infty$ and the aim
of the model is to infer the number density of loops of
a given invariant size.

Kibble \cite{kibble_evolution_1985} 
developed a model for the network evolution for the
long strings and loops in cosmology. It accounted for the
stretching of strings and collisional intercommutation (long
string segments that break off and form loops; loops that
reconnect to long string segments). A variety of models of
differing degrees of realism have been studied since then,
guided by ever more realistic numerical simulations of the
network. As a simple approximate description we focus on the
Velocity One Scale model \cite{martins_extended_1996} in a
recently elaborated form
\cite{kuroyanagi_forecast_2012,kuroyanagi_forecast_2013}.
The reattachment of loops to the
network turns out to be a rather small effect and is
ignored. We extended existing treatments by numerically
evaluating the total loop creation rate in flat
$\Lambda$CDM cosmology.  The loop energy in a
comoving volume varies like ${\dot E}_\linvariant = C \rho_\infty p v
a^3/L$ where $C$ is the chopping efficiency, $p$ is the
intercommutation probability and $v$ is the string
velocity. All quantities on the right hand side except $p$
vary in time; $C$ is a parameterized fit
in matter and radiation eras.

\begin{figure}[t]
\begin{center}
\includegraphics[width=1\linewidth]{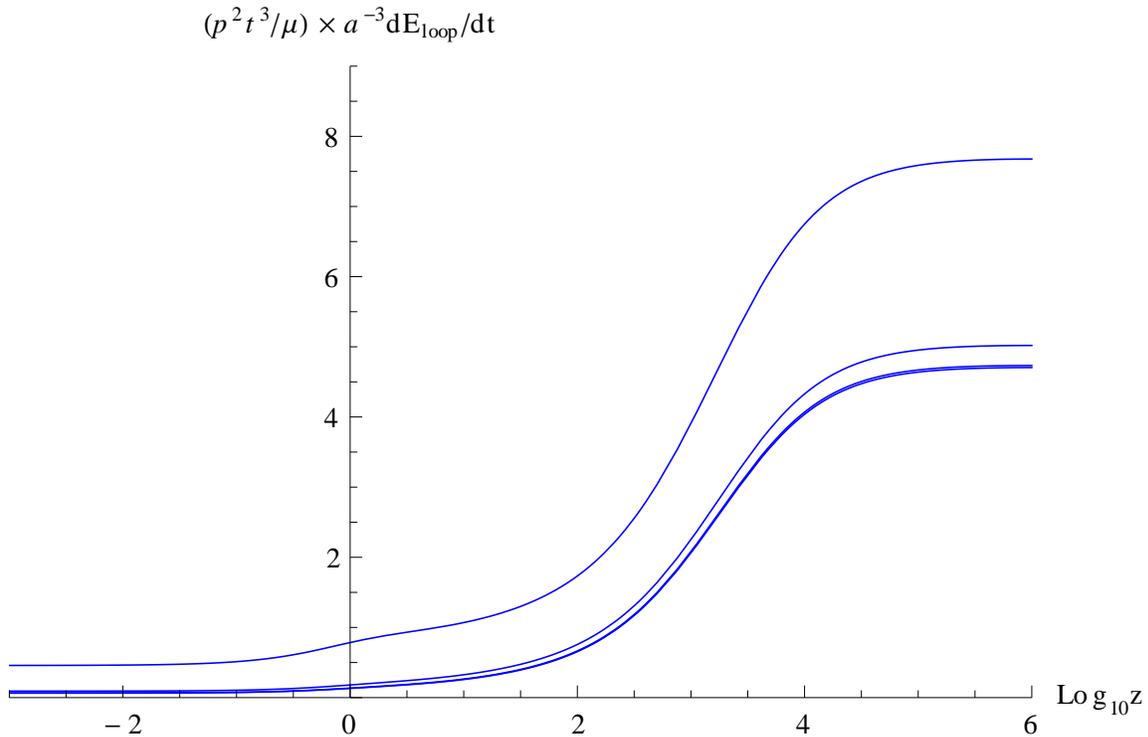}
\caption{\label{fig:edot1list} 
The redshift-dependent rate of energy loss to loops experienced by the
string network in $\Lambda$CDM cosmology for a given intercommutation
probability $p$. This is important input for predicting properties of today's loops. The ordinate is the dimensionless measure of the rate of energy loss to loops, ${\cal A} = {\dot E}_\linvariant/ \left( \mu a^3/(p^2 t^3) \right)$. If the scale factor $a(t)$ were a powerlaw in time then
${\cal A}$ would be constant with respect to $z$. This is
not the case in the $\Lambda$CDM cosmology with transitions from radiation-to-matter-to-$\Lambda$ dominated regimes. Different lines show results for
$p=1$ (top) to $p=10^{-3}$ (bottom) in powers of 10.
If network density varied as $1/p^2$ then all these lines would
overlap. This is approximately true for small $p$ but not for $p \ge 0.1$.}
\end{center}
\end{figure}

By integrating the model from large redshift to the current
epoch one evaluates the fraction of the network that is
lost to loop formation.
The rate at which the loop energy changes is ${\dot E}_\linvariant
= {\cal A} \mu a^3/(p^2 t^3)$ where ${\cal A}$ is a slowly
varying function of redshift $z$ and $p$ shown in 
Figure \ref{fig:edot1list}. The plotted variation of ${\cal A}$
with redshift indicates departures from a pure scaling solution
(consequences of the radiation to matter transition for $a(t)$ and
the implicit variation of chopping efficiency $C$). Knowing ${\cal A}$
as a function of redshift and $p$ is the input needed to 
evaluate of the loop formation rate.
Loops form with a range of sizes at each epoch.
Let us call the size at the time of formation
$\alpha t$. A very important feature predicted by
theoretical analyses is that a large fraction  of the
invariant length goes into loops with $\alpha \sim G \mu$ or
smaller. The limiting behavior of numerical simulations
as larger and larger spacetime volumes are modeled
suggests $\sim 80$\% of the chopped up long string bears
that fate. Such loops evaporate rapidly without
contributing to the long-lived local loop population. The
remaining $20$\% goes into large loops with $\alpha \sim 0.1$. 
Let us call the fraction of large loops $f=0.2$. The birth rate
density for loops born with size $\alpha \tinvariantborn$ is
\be
\left( \dndldtfrac \right) = \frac{f {\cal A}}{
\alpha p^2 \tinvariantborn^4} \delta\left( \linvariant - \alpha \tinvariantborn \right)
\ee
A loop formed at $\tinvariantborn$ with length $\linvariantborn$ shrinks by
gravitational wave emission. Its size is
\be
\linvariant = \linvariantborn - \Gamma G \mu \left( t - \tinvariantborn \right)
\ee
at time $t$ ($\Gamma \sim 50$). 

The number density of loops of size $\linvariant$ at time $t$ is the integral
of the birth rate density over loops of all length 
created in the past. For $\linvariant < \alpha t$ or,
equivalently, $\tinvariantborn < t$ we have
\ba
\dndlfrac \left(\linvariant,t\right) 
& = & \left( \frac{f {\cal A} \alpha^2 }{p^2} \right)
\left( \frac{a(\tinvariantborn)}{a} \right)^3
\frac{ \Phi^3}{\left( \linvariant + \Gamma G \mu t \right)^4} \\
\tinvariantborn & = & \frac{\linvariant + \Gamma G \mu t}{\alpha \Phi} \\
\Phi & = & 1 + \frac{\Gamma G \mu}{\alpha} .
\ea
The scaling $\dndl \propto ( \linvariant + \Gamma G \mu t)^{-4}$
was already noted by Kibble \cite{kibble_evolution_1985}.

The loop number distribution peaks at zero length but the quantity
of interest in lensing is number weighted
by loop length, $\linvariant \dndl$. The
characteristic scale at time $t$ is $\lgravcutoff =\Gamma G \mu t$,
i.e. roughly the size of a newly born loop that would
evaporate in total time $t$. The distribution $\linvariant \dndl$
peaks at $\linvariant = (2/3) \lgravcutoff$.

For $G \mu < 7 \times 10^{-9} (\alpha/0.1)(50/\Gamma)$ the loops
near $\lgravcutoff$ today were born before equipartition, $t_{eq}$. 
We use $a \propto t^{1/2}$ to simplify the expression to give
\be
\linvariant \dndlfrac = \frac{x}{\left(1+x\right)^{5/2}}
\left( \frac{ f {\cal A} }{p^2 t_0^3} \right)
\left( \Gamma G \mu \right)^{-3/2}
\left( \frac{\alpha t_{eq}}{t_0} \right)^{1/2}
\ee
where $t_0$ is today and $x \equiv \linvariant/\lgravcutoff$.

The numerical results are ${\cal A} \sim 7.68$ for
$p=1$, $f=0.2$, $\alpha=0.1$ and $\Gamma=50$ (and from $\Lambda$CDM
$t_{eq}=4.7 \times 10^4$ yr and $t_0=4.25 \times 10^{17}$ s). These give
\ba
\linvariant \left( \dndlfrac \right)_{baseline} & = & 1.15 \times 10^{-6} \frac{x}{(1+x)^{5/2}} \mu_{-13}^{-3/2} \ {\rm kpc}^{-3}\\
\lgravcutoff & = & 0.0206 \mu_{-13} \ {\rm pc}\\
\mgravcutoff & = & 0.043 \mu_{-13}^2 \ \msun .
\ea
Here, $\mu_{-13} \equiv G\mu/c^2/10^{-13}$ is an abbreviation for
the dimensionless string tension in units of $10^{-13}$.
The baseline distribution lies below
\cite{kuroyanagi_forecast_2012,kuroyanagi_forecast_2013}
on account of loss of a significant fraction of the
network invariant length to small strings and of adoption
the $\Lambda$CDM cosmology.

Next, string theory modifications to the baseline are lumped into
a common factor ${\cal G}$
\be
\left( \dndlfrac \right)_{homog} = {\cal G} \left( \dndlfrac \right)_{baseline}
\ee
to give the description of the actual homogeneous loop distribution.
Prominent among expected modifications
is the intercommutation factor $p$.  
Large scale string simulations have not reached a full understanding of the
impact of $p<1$ though there is no question that
${\cal G}$ increases as a result. The numerical
treatment of the Velocity One Scale model
implies that ${\cal A}$ is a weak function of $p$ for small $p$ and,
in that limit,
$\dndl \propto 1/p^2$. Ultimately, this matter will be 
fully settled via network simulations with $p<1$. String theory calculations
of the intercommutation probability suggests
$p=10^{-1} \text{--} 10^{-3}$ implying
${\cal G} = 10^2 \text{--} 10^6$.  

The number of populated, non-interacting throats that
contain other types of superstrings is a known unknown and
is unexplored. In our opinion, there could easily be 100's
of such throats for the complicated bulk spaces of interest.

In the string theory scenarios ${\cal G}=1$ is a
lower limit. {\it For the purposes of numerical estimates in this
paper we adopt ${\cal G} = 10^2$ (with a given tension) as the most reasonable lower
limit.} Much larger ${\cal G}$ are
not improbable while lower ${\cal G}$ are unlikely. We should emphasize that strings in different throats would have different tensions, so adopting a single tension here yields only a crude estimate.

\subsection{Local string distribution}

If a loop is formed at time $t$ with length $\linvariant =
\alpha t$ then its evaporation time $\tau =
\linvariant/\Gamma G \mu$. For Hubble constant $H$ at $t$ the
dimensionless combination $H \tau = \alpha/(\Gamma G \mu)$
is a measure of lifetime in terms of the universe's age.
Superstring loops with $\alpha=0.1$ and small $\mu$ live
many characteristic Hubble times.

New loops are born with relativistic velocity.  The peculiar
center of mass motion is damped by the universe's expansion.
A detailed study of the competing effects (formation time,
damping, evaporation, efficacy of anisotropic emission of
gravitational radiation) in the context of a simple formation
model for the galaxy shows that loops accrete when $\mu$ is
small. The degree of loop clustering relative to dark matter
clustering is a function of $\mu$ and approximately independent
of $\linvariant$. Smaller $\mu$ means older, more slowly moving
loops and hence more clustering. Below we give a simple fit to the
numerical simulations to quantify this effect.

The spatially dependent dark matter enhancement in the galaxy is
\be
{\cal E}({\vec r}) = \frac{\rho_{DM}({\vec r})}{\Omega_{DM} \rho_c}
\ee
where $\rho_{DM}({\vec r})$ is the local galactic dark matter density and $\Omega_{DM} \rho_c$ is average dark matter density in the universe. 
Low tension string loops track the dark matter with a certain
efficiency as the dark matter
forms gravitationally bound structures \cite{chernoff_clustering_2009}.
We fit the numerical results by writing
the spatially dependent string enhancement to the homogeneous
distribution as equal to the dark matter enhancement times an
efficiency factor
\ba
{\cal F}({\vec r}) & = &{\cal E}({\vec r}) \ 10^{f(y)} \\
f(y) & = & \left\{
\begin{tabular}{ll}
$-0.337 - 0.064 y^2 - 5(1 + \tanh(y-7))$ & {\rm for} $0 \le y$ \\
$-0.337$ & {\rm for} $y < 0$
\end{tabular}
\right. \\
y & = & 2 + \log_{10} \mu_{-13} 
\ea
and $\mu_{-13}$ is the dimensionless tension in units of $10^{-13}$.
The clustering is never 100\% effective because the
string loops eventually evaporate. The efficiency
saturates at $y=0$ or $G\mu/c^2 = 10^{-15}$ .
The local string population is enhanced by the factor ${\cal F}$ with
respect to the homogeneous distribution
\be
\left( \dndlfrac \right)_{local}({\vec r}) = {\cal F} ({\vec r}) \left( \dndlfrac \right)_{homog} = {\cal F} ({\vec r}) {\cal G} \left( \dndlfrac \right)_{baseline} .
\ee

To complete the description of the local loop population we
adopt empirical fits to the galaxy's dark matter halo 
\cite{binney_galactic_2008}.
Model I is a ``cored galaxy'' center in which the
central, limiting dark matter density is zero. Since
winds from stars and ejection/heating by supernovae
lift baryons out of star-forming regions,
the center, they reduce the gravitational
potential and tend to lower the dark matter density.
Model II is a ``cusped galaxy'' center in
which the density formally diverges. Many N-body simulations
show that collisionless structure formation in $\Lambda$CDM
yields such profiles. The two models are thought to bracket
the range of physical possibilities in the central
regions and are in general agreement on large scales.

\begin{figure}
\includegraphics[height=3in]{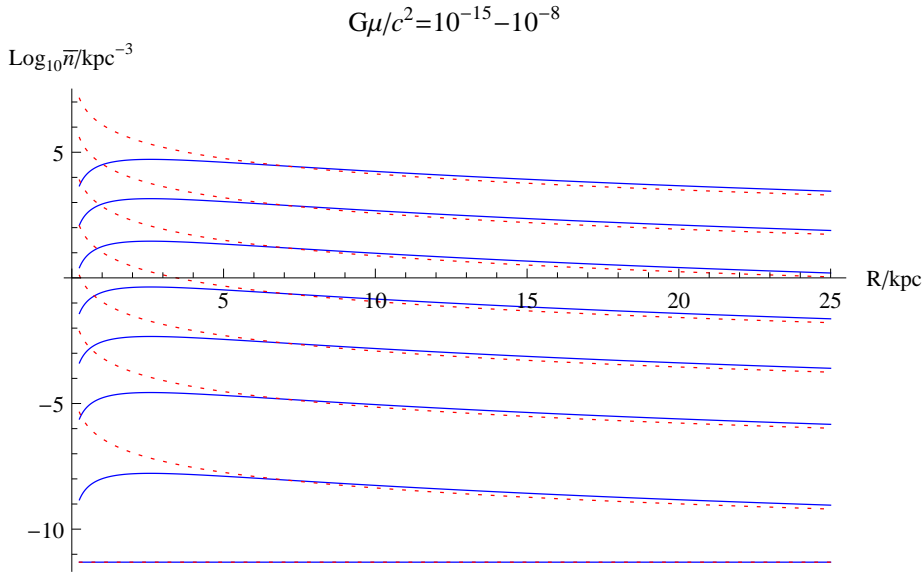}
\caption{\label{number-density-of-strings} The effective number density of
string loops in two galactic models for a range of string
tensions. The dotted red lines show results for a galactic model with
a cusp at the center, the solid blue lines for a model with a core.
The tension is $G\mu/c^2=10^{-15}$ for the uppermost
curve and increases by powers of 10 until reaching $10^{-8}$
for the lowest curve. The effective number density depends upon
the clustering of loops, the dark matter distribution within the galaxy,
the network scaling and loop distribution and ${\cal G}$ the enhancement
of string theory strings compared to field theory strings.}
\end{figure}

Let us define the effective number density of loops
as the energy density in the distribution of loops
of all size, $\rho_{loop}$, divided by energy of a loop of size equal to
the characteristic gravitational cutoff
\ba
{\bar n} & = & \frac{\rho_{loop}}{\mu \lgravcutoff} \\
         & = & \int d\linvariant \frac{dn}{d\linvariant} \frac{\linvariant}{\lgravcutoff} .
\ea
Figure \ref{number-density-of-strings} 
plots the effective number density of string loops for two
descriptions of the galaxy's dark matter distribution (core and cusp)
as a function of galactocentric radius. The different lines show a
range of string tensions, all for ${\cal G} = 10^2$.
The quantity ${\bar n}$ depends upon
the efficacy of loop clustering, the dark matter distribution within
our galaxy and the network scaling model. There are two curves 
plotted for each tension. These
agree at large radii but differ near the galactic center where
the detailed form of the dark matter distribution is uncertain.

For many experiments the rate of detection scales as
${\bar n}$ weighted by powers of $\lgravcutoff$. The microlensing rate,
for example, is proportional to the product of two factors:
the number of string loops along a given line of sight 
$\propto {\bar n}$ and the length of an individual string loop 
$\propto \lgravcutoff$.
The event rate scales like ${\bar n} \lgravcutoff \propto {\bar n} G \mu$. 
At the Sun's position, the effective number density
increase by $\sim 15$ orders of magnitude as $G \mu/c^2$ drops
from $10^{-8}$ to $10^{-15}$. The microlensing event rate increases by $\sim 8$ 
orders of magnitude. As we will discuss shortly, for the same
variation in tension the timescale for an individual microlensing event
decreases by $\sim 8$ orders of magnitude. This creates a huge
range of timescales of interest for the duration and rate of microlensing
events. 

\pad

\section{Detection}

\subsection{Detection via Microlensing}

\begin{figure}
\includegraphics[width=1\linewidth]{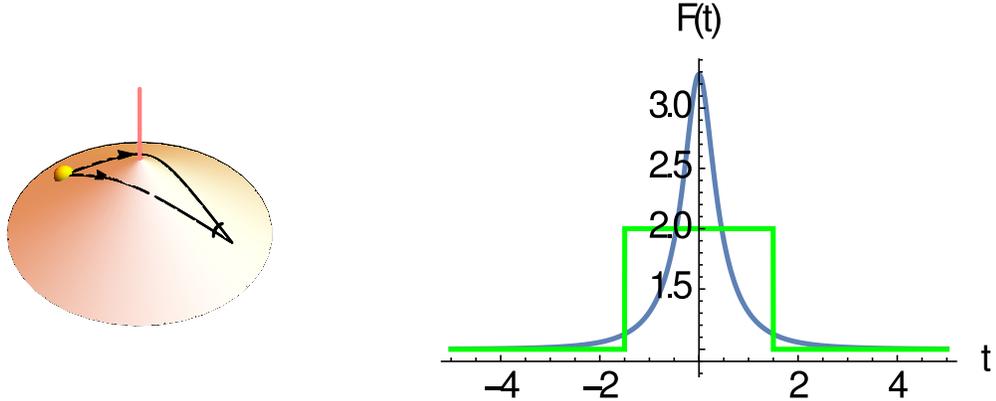}
\caption{\label{microlensing-geometry} (Left) The conical
space-time geometry near a short segment of string (red line)
permits photons to travel two paths (black arrows) from 
source (yellow ball) to observer when source, string and
observer are nearly aligned. The ideal 
observer perceives two images
of the source split by a small 
angle (black arc) proportional in size
to string tension. For low string tension, the observer cannot
resolve the separate images; however, the flux may
be easily measured to be
twice that of a single image (in the point source limit)
at each wavelength (negligible Kaiser-Stebbins effect). 
(Right) The microlensing amplification of
flux for a point source 
as a function of time for a string (green) 
is compared to a Newtonian point masses (blue; the ratio of impact parameter to Einstein radius is $0.31$). No other known astrophysical sources
produce this {\it digital} microlensing.
}
\end{figure}

Consider a straight segment of string
oriented perpendicular to the observer's line of sight
with respect to a background source as shown in
Figure \ref{microlensing-geometry}. Let two photons
from the source travel towards the string. The photons
do not suffer any relative deflection during the fly-by
as long as they pass around the string in the same
sense.  However, there is a small angular region about
the string with two paths from the source to the
observer. Background sources within angle $\Theta_E = 8
\pi G \mu = 1.04 \times 10^{-3} (G \mu/2 \times
10^{-10})$ arcsec form two images. Unlike the case of a
point mass, shear and distortion are absent. The
angular size of a sun-like star at distance $R$ is
$\Theta_\odot/\Theta_E = 4.5 \times 10^{-4} (2 \times
10^{-10}/G \mu) (10 {\rm kpc}/R)$ so galactic stellar
sources generally appear point-like for $G \mu \gta
10^{-13}$.

Compact halo objects lens background sources
\cite{liebes_gravitational_1964,paczynski_gravitational_1986} and
unresolved, lensed sources will appear to fluctuate achromatically in
brightness. This is microlensing.  Experimental efforts to detect
microlensing phenomena have borne considerable fruit
\cite{mao_astrophysical_2012}.

Likewise, loops of superstring microlens background sources but
have a special property: the
source brightness varies by a factor of 2 as the
angular region associated with the string passes across
the observer-source line of sight. The amplitude variation, schematically
compared in figure \ref{microlensing-geometry}, is quite distinctive
for point sources.
The internal motions
of a loop are relativistic and generally dominate the motion
of the source and the observer. Numerical calculations have
established that microlensing occurs when light passes
near a relativistic, oscillating string loop \cite{bloomfield_cosmic_2014};
the effect is not limited to a stationary string.

The total rate of lensing
$R_L$ implied for a distribution of loops $\dndl$ is
proportional to the solid angle swept out per time $c
l/\sqrt{3} r^2$ for loop $\linvariant$ at distance $r$, or $R_L =
\int d\linvariant dr r^2 \dndl (c \linvariant/\sqrt{3} r^2)$. Small
tensions give large lensing rates, $R_L \propto
1/\sqrt{G \mu}$ because the integral over $\dndl
\propto \linvariant^{-2.5}$ is dominated by $\linvariant \sim \lgravcutoff$, the
gravitational cutoff.

\begin{figure}
\includegraphics[width=1\linewidth]{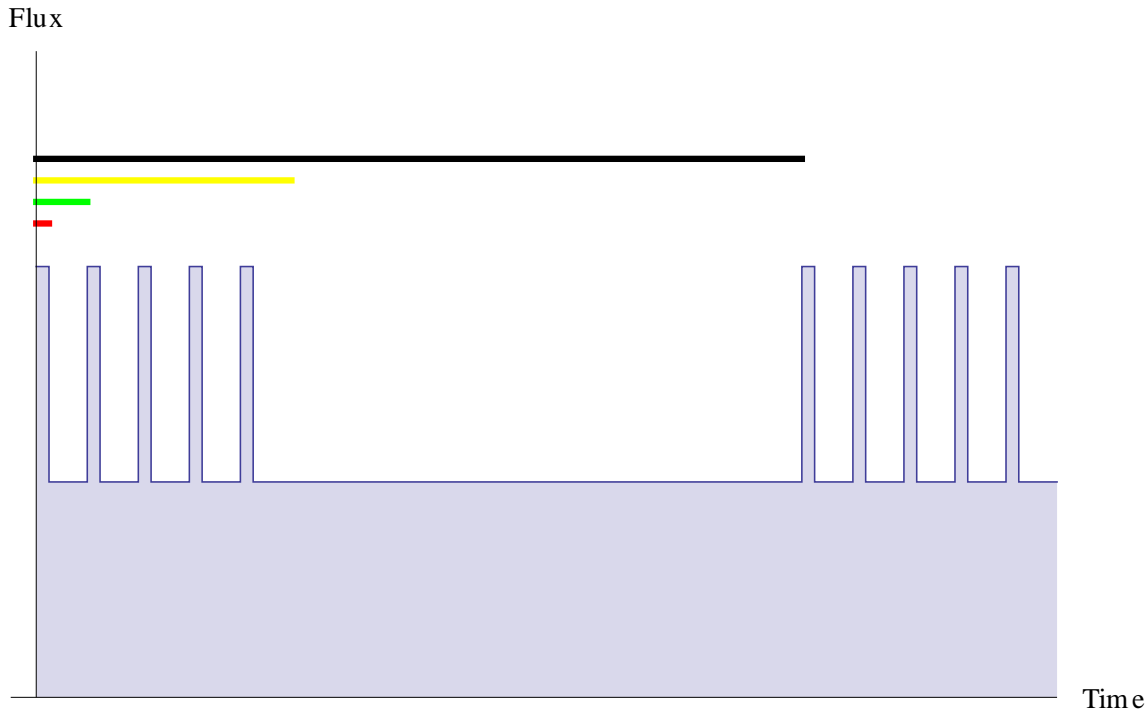}
\caption{\label{characteristic-timescales} A Schematic Pattern. Digital
microlensing doubles the flux over the time period that
source, loop and observer are aligned within the
deficit angle created by the string (red line). The
repetition interval for lensing a particular source
is the loop oscillation timescale (green
line) and $\sim 10^3$ repetitions in total occur (yellow line).
The timescale for a new source to be lensed by the
original loop (or the original source by a new loop)
is much longer (black line).}
\end{figure}

Figure \ref{characteristic-timescales} illustrates the hierarchy
of relevant timescales.
The characteristic duration of the event is $\delta t = R \Theta_E/c
\sim 630 {\rm \ s \ } (R/10 {\rm \ kpc \ })(G \mu/2 \times 10^{-10})$.  Loops
bound to the galaxy have center-of-mass motions $v_h \sim 220$ km
s$^{-1}$; microlensing of a given source will repeat $\sim c/v_h$
times; {\it new} sources are lensed at rate $\sim (v_h/c) R_L$.  The
characteristic loop oscillation timescale which governs the intervals
in repetitive microlensing of a single source is $t_{osc} \sim \lgravcutoff/c =
135 {\rm \ yr \ } (G \mu/2 \times 10^{-10})$.

The unique fingerprint of loop lensing is {\it repeated
  achromatic flux doubling} (digital microlensing).
Detection efficiency for digital lensing depends upon
the string tension, the magnitude and angular size of
the background stars and the time sampling of the
observations.  Estimates can be made for any experiment
which repeatedly measures the flux of stellar
sources. 

The first string microlensing search \cite{pshirkov_local_2010}
was recently completed using photometric data from space-based missions
CoRoT\footnote{The 2006 mission was
developed and operated by the CNES, with the contribution of
Austria, Belgium, Brazil, ESA (RSSD and Science Program), Germany,
and Spain.}\cite{corot_webpage} and RXTE\footnote{The 1995 mission was developed and operated by NASA.}\cite{rxte_webpage}. The methodology was
potentially capable of detecting strings with
tensions $10^{-16} < G \mu < 10^{-11}$ though the expected number of detections
was limited by the available lines of sight studied
in the course of the missions. In principle, any photometry experiment
that makes repeated flux measurements of an intrinsically 
stable astronomical source has the power to limit a combination of 
the number density of loops and string tension. 
One interesting possibility is the satellite GAIA launched by the European Space Agency in December 2013 designed for astrometry. It is monitoring each of about 1 billion stars about 70 times over a period of 5 years. 
Another is the Large Synoptic Survey Telescope (LSST)  to be sited in Chile within the decade to photograph the entire observable sky every few days.
Previous estimates for the rate of detections for GAIA \cite{chernoff_cosmic_2007} and LSST \cite{chernoff_white_2009} were encouraging. 
Now, detailed calculations for WFIRST are also available.

\subsection{WFIRST Microlensing Rates}

The Wide-Field Infrared Survey
Telescope\footnote{http://wfirst.gsfc.nasa.gov/} (WFIRST) is a NASA
space observatory designed to perform wide-field imaging and slitless
spectroscopic surveys of the near infrared sky. WFIRST will carry out
a microlensing survey program for exo-planet detection in the
direction of the galactic bulge by observing a total of 2.8 square
degrees for 1.2 years primarily in a broad long wavelength
filter (W-band $0.927-2.0 \mu$m).
Repeated, short exposures ($\sim 1$ minute) of the same
fields are the key to searching for and to monitoring the
amplification of bulge sources by star-planet systems along the
line of sight. Fortuitously, the WFIRST
experiment is also sensitive to cosmic strings.

\begin{figure}
\includegraphics[width=1\linewidth]{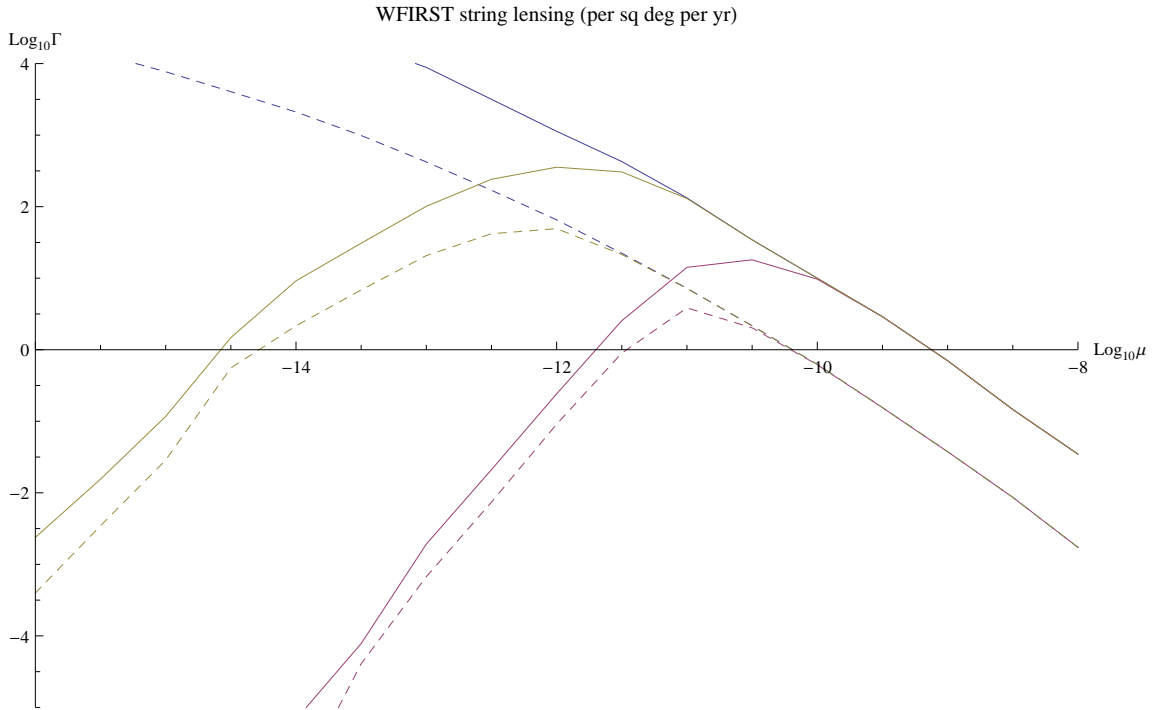}
\caption{\label{WFIRST-microlensing}
The intrinsic string lensing rate (events per sq degree per year)
for WFIRST's study of the bulge for ${\cal G}=10^2$.
The three solid lines show digital (red), analog (green) and total (blue)
event rates for the dark matter model with a cusp. The three dashed lines
(same meaning for the colors) give the event rate for the model with a core.
The analog rate does not account for blending which may diminish the
detectable rate by no more than a factor of 10.}
\end{figure}

We have evaluated\footnote{Material in this section\cite{chernoff_rates_2014}} the expected microlensing
rate by cosmic strings for a realistic distribution of stars
(stellar types, distances, velocities, etc.), dust obscuration and
survey parameters (flux sensitivity, time of exposure and angular
scale of stars). The lensing rate is split into digital events (the flux 
doubles), analog events (all potentially measurable flux enhancements
given the signal-to-noise of the observations) 
and total events (all geometric configurations that can lens,
whether detectable or not) evaluated for
two galactic dark matter models (with cusp and with core at galactic
center). Figure \ref{WFIRST-microlensing} shows
the string lensing rate per square degree
as a function of string tension $G\mu/c^2$ for ${\cal G}=10^2$.
The lensing rate for digital and analog events exceeds 10 per square
degree per year for
$10^{-14} < G\mu/c^2 < 10^{-10}$ for the cusp model and for
$10^{-13} < G\mu/c^2 < 10^{-11}$ for the core model.  
The results for the cusp and core models should
be regarded as establishing the range of astrophysical possibilities for
the loop density distribution within
the experiment. The range of string theory possibilities (the
variation of ${\cal G}$) remains substantial. Digital and
{\it spatially resolved} analog events can be reliably detected.
Blending of sources in analog
events needs to be simulated to fully characterize the fraction
of such events that will be detectable.  In any case, 
the decrease from blending is limited:
sources brighter than magnitude $23$ in the 
J-band ($1.131-1.454 \mu$m) contribute at least 10\% of the total analog
rate, are well-separated and produce light curves
detectable at high signal-to-noise.

These results show that WFIRST microlensing searches can
probe hitherto unexplored ranges of string tension. Future
surveys of wider areas
of the bulge and/or surveys lasting for longer periods of time can potentially
scale up the total expected rate of detection by
as much as 100. Surveys with shorter exposures times are able to
discern larger numbers of digital events.

\subsection{Gravitational Waves}

Besides uncovering a relic from the earliest moments of
the universe's formation, the observation of a
superstring microlensing event provides some immediate
information.  If the event is resolved in time, the
characteristic {\it string tension} is inferred.  If repetitions are seen,
the characteristic {\it loop size} is constrained. And any detection
provides a precise sky location for follow up. Some classes of
strings create electromagnetic signatures but all
generate gravitational radiation possibly observable by
LISA-like (hereafter, LISA) and LIGO-like instruments.

The scenario that LISA detects the distinctive signature of
local loop emission has been contemplated for a blind,
all-sky search \cite{depies_harmonic_2009}. They relied on
the enhancement from local clustering 
\cite{chernoff_cosmic_2007} and concluded that strings with
$10^{-19} < G \mu < 10^{-11}$ were potentially detectable
via fundamental and low-order harmonics. They also estimated
the background from the galaxy and from the universe as a
whole.

Therefore, it is no surprise that a loop microlensing
event offers some exciting possibilities.
The key considerations are: (1) The precise location
afforded by microlensing dramatically reduces the
signal search space on the sky. (2) A generic loop
radiates a full range of harmonics with frequency $f_n
= n f_1 = 2cn/\linvariant$ because the equations of motion are
intrinsically non-linear. (3) Many string
harmonics fall near LISA's peak sensitivity $n_* =
f_*/f_1 >> 1$ ($h_{LISA} \sim 10^{-23.8}$ at $f_* \sim
10^{-2.3}$ Hz for 1 year periodic variation \cite{danzmann_mission_2011};
tension $\mu_{crit} = 2.9 \times 10^{-12}$ corresponds to a 1 year period;
$n_* = 1.6 \times 10^{5} (\mu/\mu_{crit})$). 
(4) Galactic binary interference becomes problematic only
at $f < f_{WD}$ where $f_{WD} \sim 10^{-2.6}$ Hz \cite{farmer_gravitational_2003}.

LISA's verification binaries have known positions on the
sky, orbital frequencies, masses, distance limits
etc. determined by optical measurements and other means
\cite{danzmann_mission_2011}.  A long-lived superstring loop
detected by microlensing shares similar observational
advantages: it has known
position on the sky, distance upper limit estimated based on
the microlensed star and emits a distinctive signal. This creates the
ideal observational situation in which much of the data
analysis can be conducted with a single fast Fourier
transform.

Different types of loops generate different gravitational wave
signatures.  The solution for a Nambu-Goto loop's motion is the sum of
left and right-moving one-dimensional waves subject to nonlinear
constraints. When both modes are smooth (no derivative
discontinuities) the dynamical solution can form a cusp. At that
instant a bit of the string reaches the speed of light.  When one mode
is smooth but the other has a discontinuity the loop's dynamical
solution includes a kink, a discontinuity in the string's tangent
vector that moves along the string at the speed of light in one
direction.  When both modes have discontinuities we call the solution
a k-kink. The presence of cusps, kinks and k-kinks may be
inferred from the paths of the tangent vectors of the left and right
moving modes which are constrained to lie on the surface of a sphere.
These distinctions are relevant for gravitational wave
emission: the cusp generates bursts of beamed
radiation, the kink has a pulsar-like beam that generically
sweeps across a substantial angle in the sky and the k-kink
emits into a substantial fraction of the sky.

\def\gwextradimensional{
2011JCAP...03..004O,
2010JCAP...09..013O,
2010PhRvL.105h1602O}

The tendency for string loops to explore part of the internal space of
a typical compactification during oscillations
\cite{avgoustidis_cosmic_2012} alters the cusp and kink beaming
patterns with potentially important repurscussions for rates of
detection in gravitational wave experiments\cite{\gwextradimensional}.

\begin{figure}
\includegraphics[height=3in]{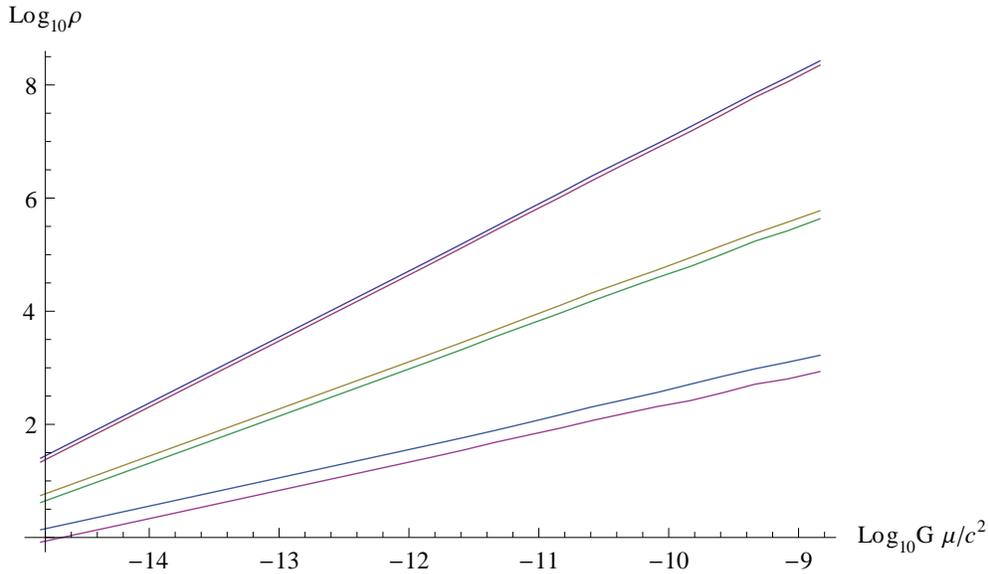}
\caption{\label{detection} The log of LISA's signal strength
  ($\log_{10} \rho_{SNR}$) for 1 year of observation of
  cusp, kink and k-kink (top to bottom) at a distance of 1
  kpc with a known position on the sky as a function of
  string tension. The orientation of the cusp and kink is
  for the maximum signal. Upper/lower line pairs show the
  effect of ignoring/accounting for white dwarf
  interference. For tensions with $G \mu/c^2 > 3 \times
  10^{-12}$ the fundamental loop period exceeds the length
  of the observation so excess power is seen but the
  frequency resolution is insufficient to measure individual
  harmonics.}
\end{figure}

We model the beaming anisotropy, the harmonic power and phase of the
gravitational wave signal (qualitatively and quantitatively
extending previous analyses \cite{depies_harmonic_2009}).
Standard signal detection theory characterizes the strength
of detection and the precision with which measurements can
be made by comparing the putative signal
to noise sources. Here, we consider only detector noise
described as an additive, stationary, random Gaussian process with
one-sided spectral noise at frequency $f$ is
$S( \| f \| )$. The strength of signal $h(t)$ is
$\rho_{SNR} = \sqrt{2 \left< h, h \right>}$ where
the symmetric inner product is $\left< g, h \right> =
\int_{-\infty}^{\infty} df \frac{{\tilde g}(f) {\tilde
    h}^*(f)}{S( \| f \| )} $ and ${\tilde g}(f)$ is the Fourier
transform of $g(t)$ (the factor of $2$ stems from
one-sided noise versus two-sided signal).  Figure
\ref{detection} summarizes $\rho_{SNR}$ for detecting
a single source with beamed cusp, beamed kink and generic,
unbeamed emission. Upper and lower line pairs show the effect
of ignoring and accounting for the galactic white dwarf interference. Confusion
is minimal because the string signal extends to
frequencies where the binary density is small and because
the signal overlap between string and binary sources is
small.

The time-averaged microlensing rate is roughly proportional
to the loop's maximum projected area divided by its period
of oscillation but the gravitational wave emissivity is
dominated by specific loop configurations having rapidly
moving segments when the loop is most
contracted. Microlensing and beamed, gravitational wave
emission generally will not and need not be
simultaneous to be informative. We elaborate on this
important point.

Generic cusps and kinks have characteristic beaming patterns
described above and undergo fully three dimensional motions
that yield time-averaged microlensing rates that are not
particularly angle-dependent. Microlensing is a favorable
means to locate such loops but the high frequency, directed
gravitational wave beams they emit are visible only from
special directions. A microlensing detection does not
significantly influence the probability of intercepting
beamed emission in experiments with duration exceeding the
fundamental loop period. Instead,
there is a roughly constant, low probability that a
cusp that is present will beam in the observer's direction and modest
probability that a kink will (the beam's angular size is smaller at higher
frequencies of emission).

The most likely configuration for k-kinks is a flat, degenerate
box-like orbit with a well-defined plane. Special observers
who lie in that plane will not see microlensing but most
others are sensitive to the loop's projected motion and
see the effect if a suitable background source exists. The
gravitational wave emission of galactic k-kinks is not strongly
beamed but, as the signal estimates show, is still
potentially detectable. Unbeamed emission from loops
that happen to contain cusps and kinks should also be
detectable whether or not the high frequency beam
impinges on the observer.

To summarize, there are excellent chances to detect some of
this emission once the precise location of the source is
known. If the fundamental string period is less than the
duration of the experiment then the string's frequency comb
can be resolved and fit.  Otherwise, there will be excess
frequency-dependent power seen in the microlensing
direction. The issue of confusion with local and
cosmological string loops will be addressed in future studies.

While non-detection by LISA would place bounds on tension
and loop size the most informative scenario will be
measuring the individual harmonics of gravitational wave
emission.  Cusps have almost smooth spectral amplitude and
phase distributions, kinks show some mode-to-mode variation
and k-kinks progressively more. The variation with harmonic
order provides immediate information regarding derivative
discontinuities, e.g. {\it an experimental determination of
  cusp and kink content}. This is directly applicable for
understanding the string network in a cosmological
context. Specifically, the discontinuities are created when
horizon-crossing strings are chopped into loops but are
modified as the strings are stretched out by the expanding
universe. It is important to understand if the loops
exclusively contain discontinuities or, also, cusps.  This
relates to anticipated cosmologically distant string sources
that contribute to the gravitational wave background
currently sought.

A measurement of the fundamental frequency is equivalent to
a precise measurement of the invariant string length. It can be
determined in 1 year to relative accuracy at least as good
as $\Delta f_1/f_1 \sim 1/\rho_{SNR}$. The gravitational
decay time for the typical string loop is likely to be of
order the age of universe. A study of Figure \ref{detection}
suggests that the expected decay in a year $\Delta f_1/f_1
\sim 10^{-10}$ is too small to measure directly unless the
loop is unexpectedly close to the end of its life or very
nearby.  Instead, first, one will measure changes in
frequency from center of mass motion and acceleration within
the galaxy and place upper limits on $\Delta f_1/f_1$. The
latter will constrain string couplings to axion-like fields
and astrophysical effects of nearby matter.
Since loops are long-lived sources emitting over a wide
range of harmonics, eventually, multiple detectors will be
brought to bear and the loop's decay measured in a direct
fashion. This program can yield a precise determination of the
string tension. 

\section{Summary}

String theory contains a consistent quantum gravity sector and
possesses a deep mathematical and physical structure. It has become
apparent that the theory's richness enables it to describe a huge
multiplicity of solutions and that testing whether string theory is {\it
  the} theory of nature is no easy task. Fortunately and
fundamentally, it possesses an unambiguous property, namely, the
presence of strings. If the evolution of the early universe is able to
produce stable or meta-stable closed strings that stretch across the sky, then
these unusual objects may provide distinctive evidence for the
theory. As more and more observational data have been collected and
analyzed, most cosmologists now conclude that the universe started
with an inflationary epoch. This conclusion quite naturally motivates
investigations of the following interrelated issues:
how inflation is realized, how and
what type of cosmic strings are produced, how to search for and
detect these objects and how to measure their properties.

Here, we review how inflation may be realized within string theory and
whether there are distinctive stringy features in string
theory-motivated inflationary scenarios. In lieu of presenting a full
review we sample some of the scenarios proposed already to give
readers a taste of what can be achieved. As there are many ways to realize inflation 
within the string theory framework,  we hope better data will eventually indicate how
 inflation happens in string theory. As we have pointed out, some stringy inflationary 
 scenarios can have interesting distinct features to be detected.
We are also encouraged to note that 
cosmic superstrings are naturally produced after inflation.

Next, we review and discuss cosmic superstring features and
production.  One underlying message is that, because of the
intricacies of flux compactification, multiple types of cosmic
superstrings with a wide range of properties may be produced.  Such
superstrings are quite different from the vortices well studied within
field theory. The tension may be drawn from a discrete spectrum and
different string states can cooperate to form junctions and beads. A
future experimental observation of any of these properties can
legitimately be taken as a smoking gun for string theory.


We review ongoing searches for strings in cosmology.  Although
specific experimental results are model-dependent the
diversity of approaches has jointly lowered the upper limit for string
tension by about two or more orders of magnitude below GUT scales. This is
important progress. However, well-motivated string models with warped geometries
can yield characteristic string tensions that are far smaller than current upper
limits, perhaps all the way to the general scales of the Standard
Model itself. As Feynman famously said about miniaturization,
``there's plenty of room at the bottom''. Likewise, for string theory,
there is a huge range of plausible tensions yet to be explored.

We review the new physics and cosmology of these relatively low
tension cosmic strings, particularly the tendency for string loops to
track structure formation in the matter-dominated eras. As low tension strings 
decay slow enough, their clustering happens like dark matter, with a $10^5$ 
order of magnitude enhancement in density within the galaxy. This
opens up intriguing possibilities to seek strings having
tensions many orders of magnitude less than current upper limits.

We review one promising possibility that, for tension $10^{-14} \le G \mu <10^{-7}$, 
string loops within the Galaxy microlens stars within the Galaxy. This type of microlensing
qualitatively  differs from the well-studied microlensing effect of compact objects
in the Galactic halo. Ongoing (e.g., GAIA of ESA) as well as planned (e.g.,
LSST, WFIRST) searches for variable stars and/or exo-planets are
sensitive to cosmic superstring microlensing.  This gives us hope that
much improved upper limits on string tension and/or actual detections
will emerge in the coming years. We outline the important role that
gravitational wave observations might play if and when a string loop
is detected.

\bigskip

We thank Jolyon Bloomfield, Tom Broadhurst, Jim Cordes, John Ellis,
Eanna Flanagan, Romain Graziani, Mark Hindmarsh, Renata Kallosh,
Andrei Linde, Liam McAllister, Levon Pogosian, Ben Shlaer, Yoske
Sumitomo, Alex Vilenkin, Barry Wardell, Ira Wasserman and Sam Wong for
discussions. We gratefully acknowledge the support of the John
Templeton Foundation (Univ. of Chicago 37426-Cornell FP050136-B).
DFC is supported by NSF Physics 
(Astrophysics and Cosmology Theory, No. 1417132);
SHHT is supported by the CRF Grants of the Government of the Hong Kong SAR under HUKST4/CRF/13G.


\bibliographystyle{JHEP.bst}
\bibliography{extracted}

\end{document}